\documentclass[]{aa}
\usepackage{graphicx}
\usepackage{color}
\usepackage{txfonts}\usepackage{subcaption}
\usepackage{natbib}
\usepackage{graphicx}
\usepackage{txfonts}
\usepackage{newtxtext,newtxmath}
\usepackage{ae,aecompl}
\usepackage{amsmath}	
\usepackage{amssymb}	
\usepackage{float}
\usepackage[]{hyperref}
\begin{document}

   \title{A broadband X-ray view of the NLSy1 1E 0754.6+392.8}

\titlerunning{The complex X-ray spectrum of 1E 0754.6+392.8}

   \author{R. Middei
          \inst{1}\fnmsep\thanks{riccardo.middei@uniroma3.it}
          \and
          F. Tombesi\inst{2,3,4,5} \and F. Vagnetti \inst{2} \and R. Serafinelli \inst{6} \and S. Bianchi\inst{1} \and \\ G. Miniutti\inst{7} \and A. Marinucci\inst{8} \and G. A. Matzeu \inst{9} \and P.-O. Petrucci \inst{10}  \and F. Ursini \inst{11}  \and A. Zaino \inst{1}
	}
\institute{Dipartimento di Matematica e Fisica, Universit\`a degli Studi Roma Tre, via della Vasca Navale 84, I-00146 Roma, Italy
\and
Dipartimento di Fisica, Universit\`a di Roma "Tor Vergata", via della Ricerca Scientifica 1, I-00133, Roma, Italy
\and
INAF Astronomical Observatory of Rome, Via Frascati 33, 00078 Monteporzio Catone, Italy
\and
Department of Astronomy, University of Maryland, College Park, MD 20742, USA
\and
NASA/Goddard Space Flight Center, Code 662, Greenbelt, MD 20771, USA
\and
INAF - Osservatorio Astronomico di Brera, Via Brera 28, 20121, Milano, Italy
\and
Centro de Astrobiolog\'ia (CSIC-INTA), Dep. de Astrof\'isica; LAEFF, Villanueva de la Ca$\tilde{n}$ada, Madrid, Spain
\and
ASI - Unit\`a di Ricerca Scientifica, Via del Politecnico snc, 00133 Rome, Italy
\and 
European Space Agency (ESA), European Space Astronomy Centre (ESAC), E-28691 Villanueva de la Ca$\tilde{n}$ada, Madrid, Spain
\and
Univ. Grenoble Alpes, CNRS, IPAG, F-38000 Grenoble, France
\and
INAF/IASF Bologna via Gobetti 101, 40129 Bologna, Italy
}


\abstract
{The soft X-ray band of many active galactic nuclei (AGN) is affected by obscuration due to partially ionised matter crossing our line of sight. In this context, two past \textit{XMM-Newton} observations (6 months apart) and a simultaneous \textit{NuSTAR}-\textit{Swift} ($\sim$8 years later) exposure of the Narrow Line Seyfert 1 galaxy 1E 0754.6+392.8 revealed an intense and variable WA and hints of additional absorbers in the Fe K$\alpha$ band.}
{We aim at providing the first X-ray characterisation of this AGN discussing its broadband (0.3-79 keV) spectrum and temporal properties.}
{We conduct a temporal and spectroscopic analysis on two $\sim$10 ks (net exposure) \textit{XMM-Newton} snapshots performed in April and October 2006. We also study the high energy behaviour of 1E 0754.6+392.8 modelling its broadband spectrum using simultaneous \textit{Swift}-\textit{NuSTAR} data. Both phenomenological and physically motivated models are tested.}
{We find the presence of flux variability ($\sim$150\% and 30\% for 0.3-2 and 2-10 keV bands, respectively) and spectral changes at months timescales ($\Delta\Gamma\sim$0.4). A reflection component that is consistent with being constant over years and arising from relatively cold material far from the central super massive black hole is detected. The main spectral feature shaping the 1E  0754.6+392.8 spectrum is a warm absorber. Such a component is persistent over the years and variability of its ionisation and column density is observed down on months in the ranges 3$\times10^{22} \rm cm^{-2}\lesssim$ N$_{\rm{H}}\lesssim7.2\times10^{22} \rm cm^{-2}$ and  1.5 $\lesssim\log(\xi/{\rm erg~s^{-1}~cm})\lesssim$2.1. Despite the short exposures, we find possible evidence of two additional high-ionisation and high-velocity outflow components in absorption.}
{Our analysis suggests the existence of a complex system of absorbers in 1E 0754.6+392.8. Longer exposures are mandatory in order to characterise, on more solid grounds, the absorbers in this AGN.}

	\keywords{galaxies:active – galaxies:Seyfert – quasars:general – X-rays:galaxies
}

\maketitle
%

\section{Introduction}

Active galactic nuclei (AGNs) lie in the central region of galaxies, and their emission is observed from gamma rays down to radio frequencies. Most of the released energy is emitted in the optical-UV band due to accretion of gas from a disc surrounding a supermassive black hole \citep[SMBH,][]{M87}. Moreover, most AGNs are luminous in the X-ray band, and this energetic emission can be explained in terms of an inverse Compton mechanism involving seed disc photons and a distribution of thermal electrons overlying the disc, the so-called hot corona \citep[e.g.][]{haar91,haar93}. In X-rays, AGNs display a power law-like spectrum \cite[e.g.][]{Guai99b,Bian09}, which depends on the physical conditions of the coronal plasma (i.e. electron temperature and optical depth). At hard X-ray energies, an exponential cut-off is often observed \citep[e.g.][]{Fabi15,Fabi17,Tort18}, and it is interpreted as a further signature of the nuclear Comptonisation \citep{Rybi79}. The coronal emission can further interact with the SMBH surroundings, and the emerging spectra can be modified by absorption and reflection. The reprocessing of the primary emission gives rise to additional spectral features such as a Compton hump at about $\sim$30 keV \citep[e.g.][]{Matt91,George91} and a fluorescence Fe K$\alpha$ emission line. \\
\indent A detailed analysis of absorption profiles in X-ray spectra can provide additional information about the surroundings of the central engine. About 50\% of AGNs \citep{Reyn97} display soft X-ray absorption features due to ionised gas along the line of sight which are indicative of a warm absorber \citep[WA,][]{Blustin05}. Such phenomenon, first reported in \citet{Halpern1984}, consists in a spectral dip, indicating distant outflowing material covering the inner X-ray emission at $\sim$1 keV. High resolution spectroscopy provides the best way to discern and characterise the outflowing components \citep[e.g.][]{Longinotti2010,Behar17,Laha2014,Laha2016,Mao2019}, though low resolution studies still provide insights and can be applied on a larger number of sources \citep[e.g.][]{Piconcelli2005,Tomb10,Goff13,Capp16}. The ionisation parameter of the WAs is typically in the range $\log(\xi/{\rm erg~s^{-1}~cm})\simeq$0--3 and the equivalent hydrogen column density is between N$_{\rm H}$$\simeq$$10^{20}$--$10^{22}$ cm$^{-2}$. The relative absorption lines and edges are often blue-shifted, indicating that the gas is outflowing with velocities from v$_{\rm out} \simeq 100$ km~s$^{-1}$ up to v$_{\rm out} \simeq 1000$ km~s$^{-1}$. \\
\indent Highly blueshifted Fe K absorption lines indicative of ultra fast outflows (UFOs) with velocities higher than 10000 km~s$^{-1}$ have been reported in the X-ray spectra of several AGNs \cite[e.g.][]{Tomb10, Goff13}. The ionisation of this outflowing plasma can be very high, in the range $\log(\xi/{\rm erg~s^{-1}~cm})\simeq$3-6 and the column density is also large, up to values of $N_H \simeq 10^{24}$~cm$^{-2}$ \cite[e.g.][]{Tomb11}. Recent studies are reporting on the presence of multi-structured disc winds. \cite{Reeves2018} found an additional component of the fast wind in PDS 456 while caught in the a low-flux state with \textit{XMM-Newton} and \textit{NuSTAR}, while the case of MCG-03-58-007 is discussed in \cite{Braito18} and \cite{Matzeu19}. Moreover, a positive correlation between the outflow velocity of the UFOs and the X-ray luminosity \citep[e.g.][]{Matzeu17,Pinto2018} has been observed, and this is expected in a radiatively driven wind scenario. 
These disc winds are observed at sub-parsec scales from the central SMBH and seem to be powerful enough to affect the host galaxy environment \cite[e.g.][]{Tomb12}. Indeed, the recent detection of UFOs in some ultra-luminous infrared galaxies (ULIRGs) showed that they are likely responsible for driving the observed massive, large-scale interstellar matter (ISM) outflows \cite[e.g.][]{Tomb15,Tomb17,Feru15,Fiore17,Veil17} , and that they may quench star formation as expected from AGNs feedback models \cite[e.g.][]{Zubo12,Fauc12}. \\
\indent In this paper we focus on the X-ray analysis of 1E 0754.6+392.8 . This object is one of the two brightest AGNs in the \textit{NuSTAR} serendipitous source catalogue \citep{Lans17}, the other being HE 0436-4717 \citep{Midd18b}. 1E 0754.6+392.8 is classified by \cite{Bert15} as a local ($z=0.096$) radio quiet Narrow Line Seyfert 1 galaxy \citep[NLS1, see also][]{enya02}. For the mass of the central black hole, we adopt the single-epoch estimate by \citet{Bert15}, $\log\,$M$_{\rm{BH}}$/M$_{\odot}=8.15$, consistent with the reverberation-based value $\log\,$M$_{\rm{BH}}/M_{\odot}=8.0$ reported by \citet{Serg07}. The bolometric luminosity and Eddington ratio are estimated as $\log\,$(L$_{\rm{bol}}/\rm erg~ s^{-1})$=45.4 and $\log\,$(L$_{\rm bol}$/L$_{\rm Edd}$)=-0.85 \citep{Bert15}. Although several works discuss the optical properties of 1E 0754.6+392.8, this source is poorly studied in the X-rays. The Einstein observatory \citep{Giacconi79} observed this AGN and \citet{Gioia90} reported its flux to be F$_{\rm{0.3-3.5~keV}}$=1.8$\times$10$^{-12}$ ergs cm$^{-2}$ s$^{-1}$. Throughout the paper, the standard cosmology \textit{$\Lambda$CDM} with H$_0$=70 km s$^{-1}$ Mpc$^{-1}$, $\Omega_m$=0.27, $\Omega_\lambda$=0.73, is adopted.

\section{Data reduction}
\begin{table}
	\centering
	\setlength{\tabcolsep}{.9 pt}
	\caption{\small{The ID, the telescope, the net exposure and the start data of the observations here analysed are reported.}}
	\begin{tabular}{c c c c c c}
		\hline
		Satellite &Instrument&Obs. ID& Net exp. &Start-date \\
		&&&(ks) &yyyy-mm-dd \\
		\hline
		\textit{XMM-Newton} &\textit{PN}&0305990101&  13.5 & 2006-04-18 \\
		\hline
		\textit{XMM-Newton} &\textit{PN}&0406740101&  14.7 &2006-10-22\\
		\hline
		\textit{NuSTAR} &\textit{FPMA/B}&60001131002 & 45.1 &2014-09-12\\
		\hline
		\textit{Swift} &\textit{XRT}&00080595001 &2.7  &2014-09-12\\
	\end{tabular}
\end{table}

\begin{figure}	
	\includegraphics[width=0.49\textwidth]{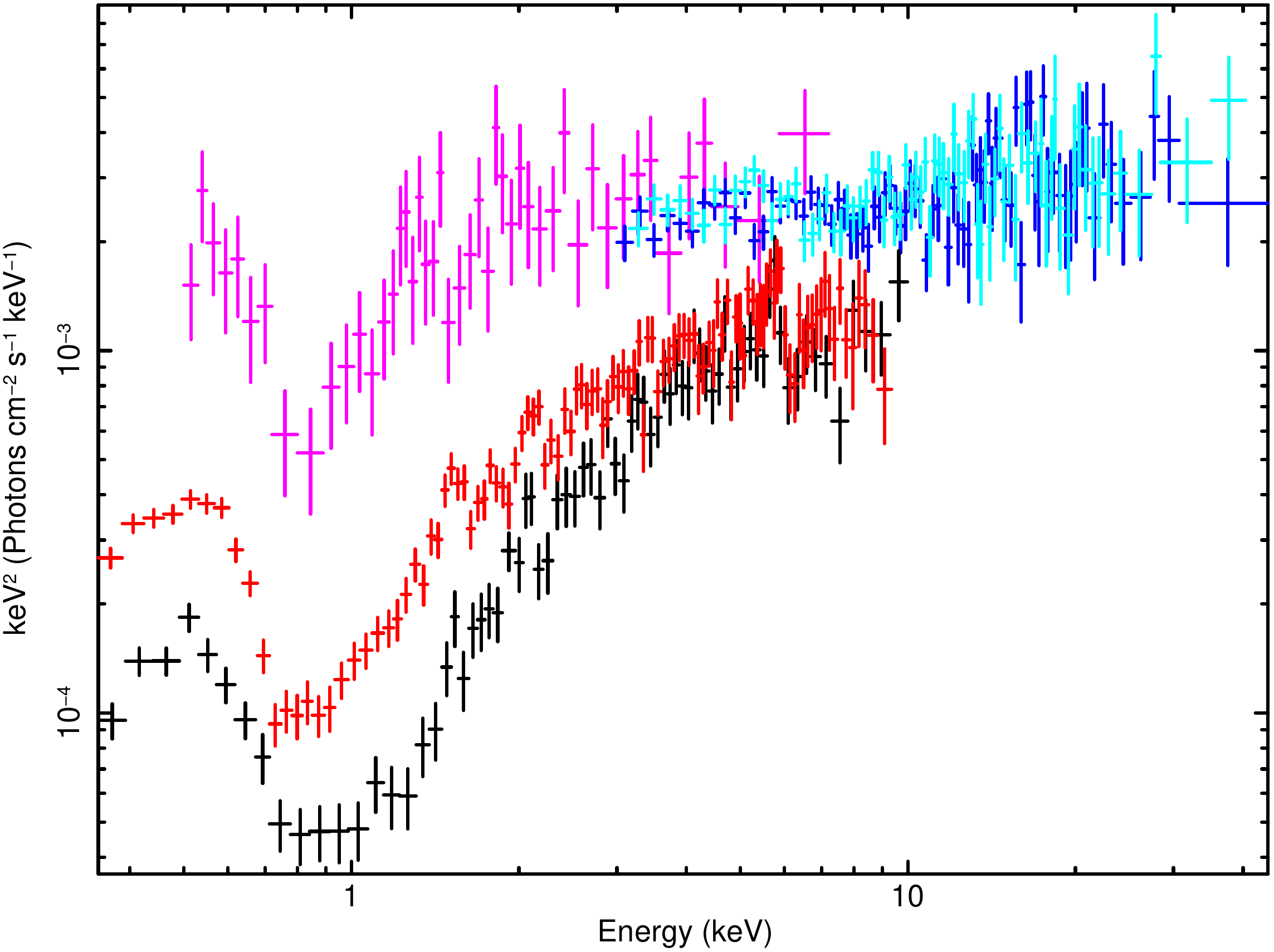}
	\caption{\small{The unfolded spectra of 1E 0754.6+392.8 as observed by \textit{XMM-Newton} (black for Obs. 1 and red for Obs. 2), \textit{Swift} (magenta) and \textit{NuSTAR} (blue and cyan). Absorption in the soft X-rays is clearly observed and it is persistent over the years. This colour code is used throughout the paper. Finally, the underlying model consists in a power-law with $\Gamma$=2 and unitary normalisation.}\label{unf}}
\end{figure}

This work takes advantage of two archival \textit{XMM-Newton} observations and one \textit{Swift-NuSTAR} simultaneous observation (see the log in Tab. 1). The \textit{XMM-Newton} \citep{Jans01} observatory pointed twice 1E 0754.6+392.8 on April 18 2006 and October 22 2006. \textit{NuSTAR} \citep{Harr13} observed the source serendipitously on September 12 2014 \citep{Lans17} simultaneously with \textit{Swift}.\\
\indent \textit{XMM-Newton} data were processed using the \textit{XMM-Newton} Science Analysis System ($SAS$, Version 18.0.0). To select the extraction radius of each observation and to screen for high background time intervals we used an iterative process that maximises the Signal-to-Noise (S/N) ratio in the 3-9 keV band \citep[details in][]{Pico04}. We therefore use a 19 arcsec radius for extracting the source in Obs. 1, while a 40 arcsec of circular region was found to maximise the S/N ratio for Obs. 2. We extract the background of both the observations from a circular region of 50 arcsec radius located on a blank area of the detector close to the source. We binned all the spectra to have at least 25 counts for each bin, and not to oversample the instrumental energy resolution by a factor larger than 3.
Data from the \textit{MOS} detectors have a much lower statistics even when the spectra are co-added. Being our analysis mainly focused on the Fe K energy band, we decided to use only \textit{PN} due to the lower statistics of the co-added \textit{MOS} spectrum.\\
\indent The \textit{NuSTAR} observation was reduced in accordance with the standard procedure described in Perri et al., 2013, and using HEASOFT (v. 6.25), NuSTARDAS (v 1.8.0) and the `x20180710' version of the calibration database. Spectra were extracted for both the hard X-ray detectors \textit{FPMA/B} on the \textit{NuSTAR} focal plane. A circular extraction region with 40 arcsec radius was used for the source, while the background was obtained adopting a region of the same size in a blank area of the same chip. The \textit{NuSTAR} spectra were binned not to over-sample the instrumental resolution by a factor larger than of 2.5 and to have a S/N greater than 3 in each spectral channel.\\
\indent Finally, we used online facilities provided by the ASI Space Science Data Center (SSDC) for processing and reducing the \textit{Swift/XRT} data (multimission archive, \url{http://www.ssdc.asi.it/mma.html}). The spectrum was extracted from a circular region with a radius of 20 arcsec centered on the source while the background was sampled from an annular region extending between 40 arcsec and 80 arcsec around the source. For spectral fitting we use the source spectrum binned to a minimum of 20 counts per bin.\\
The spectra of the whole dataset, unfolded using a $\Gamma$=2 power law model with common normalisation, are shown in Fig.~\ref{unf}.

\section{Spectral and timing analysis}

We consider the \textit{EPIC-pn} data in the E=0.3-10 keV energy band and the \textit{NuSTAR-Swift/XRT} simultaneous observation in the 0.5-79 keV band. We point out that \textit{XMM-Newton/NuSTAR} exposures are about 8 years apart, thus their corresponding spectra are not simultaneously fitted. A Galactic column density of N$_{\rm{H}}$=5.6$\times10^{20}$ cm$^{-2}$ \citep{HI4PI} is always considered when fitting the spectra. Finally, in the text errors are quoted at 90\% confidence level and errors in the plots account for 68\% uncertainties.

\subsection{\textit{Swift/XRT}-\textit{NuSTAR}: The 0.5-79 keV band}

\begin{figure}
	\includegraphics[width=0.48\textwidth]{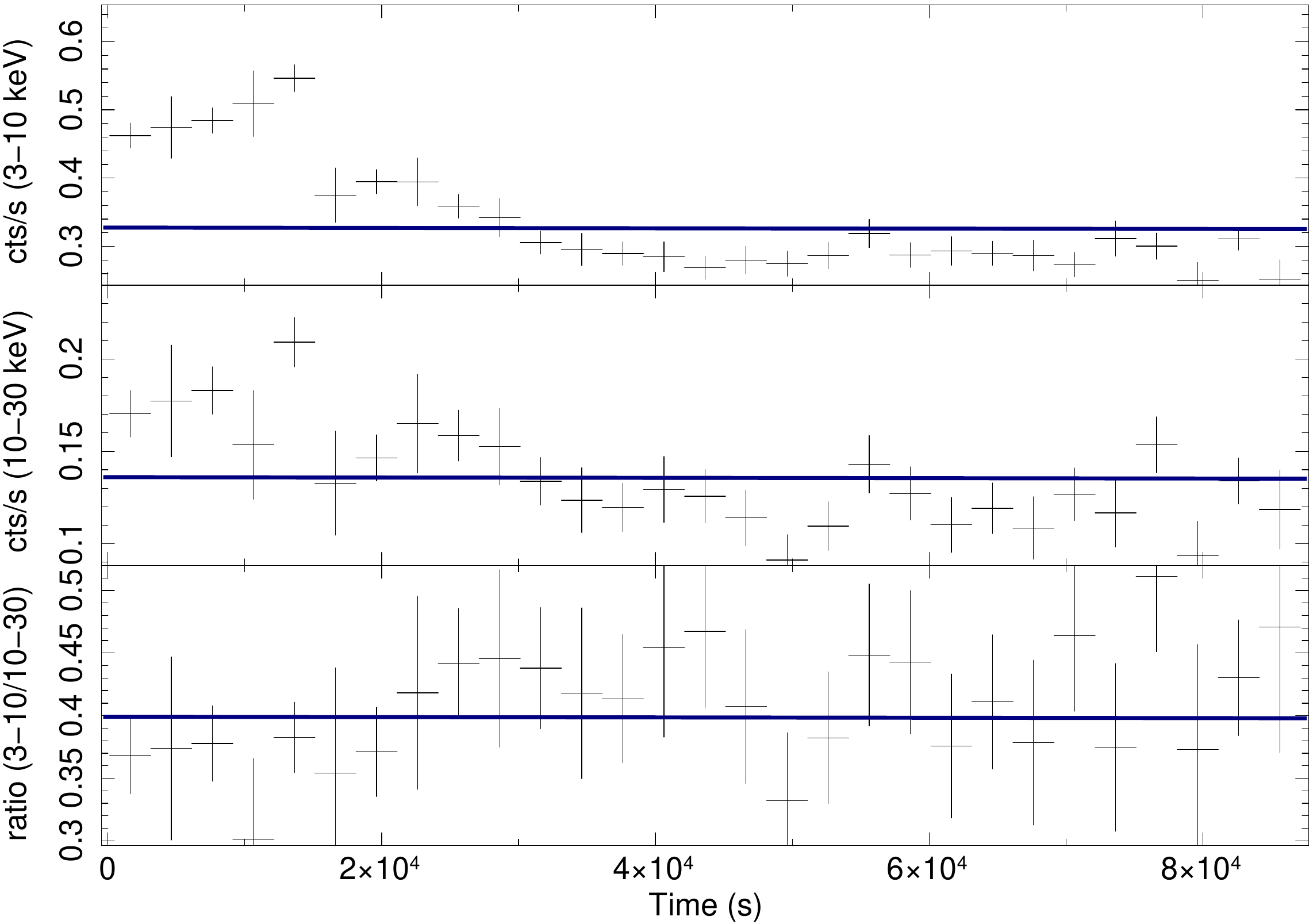}
	\caption{\small{The background subtracted \textit{NuSTAR} light curves in the 3-10 and 10-30 keV bands are showed. In the bottom panel the ratio between the two bands is displayed. The adopted time bin is 3000 s}. The blue lines account for the average counts number in the two bands and for their ratio.\label{Nulc}}
\end{figure}

We investigated the temporal properties of the \textit{NuSTAR} observation first. The 3-10 keV light curve reveals the presence of intra-observation variability (up to a factor of $\sim$2) while a more constant behaviour characterises the 10-30 keV band, see Fig.~\ref{Nulc}. No significant spectral variability is observed, thus we considered the source spectrum integrated over the whole observation length.

\begin{figure}
	\includegraphics[width=0.48\textwidth]{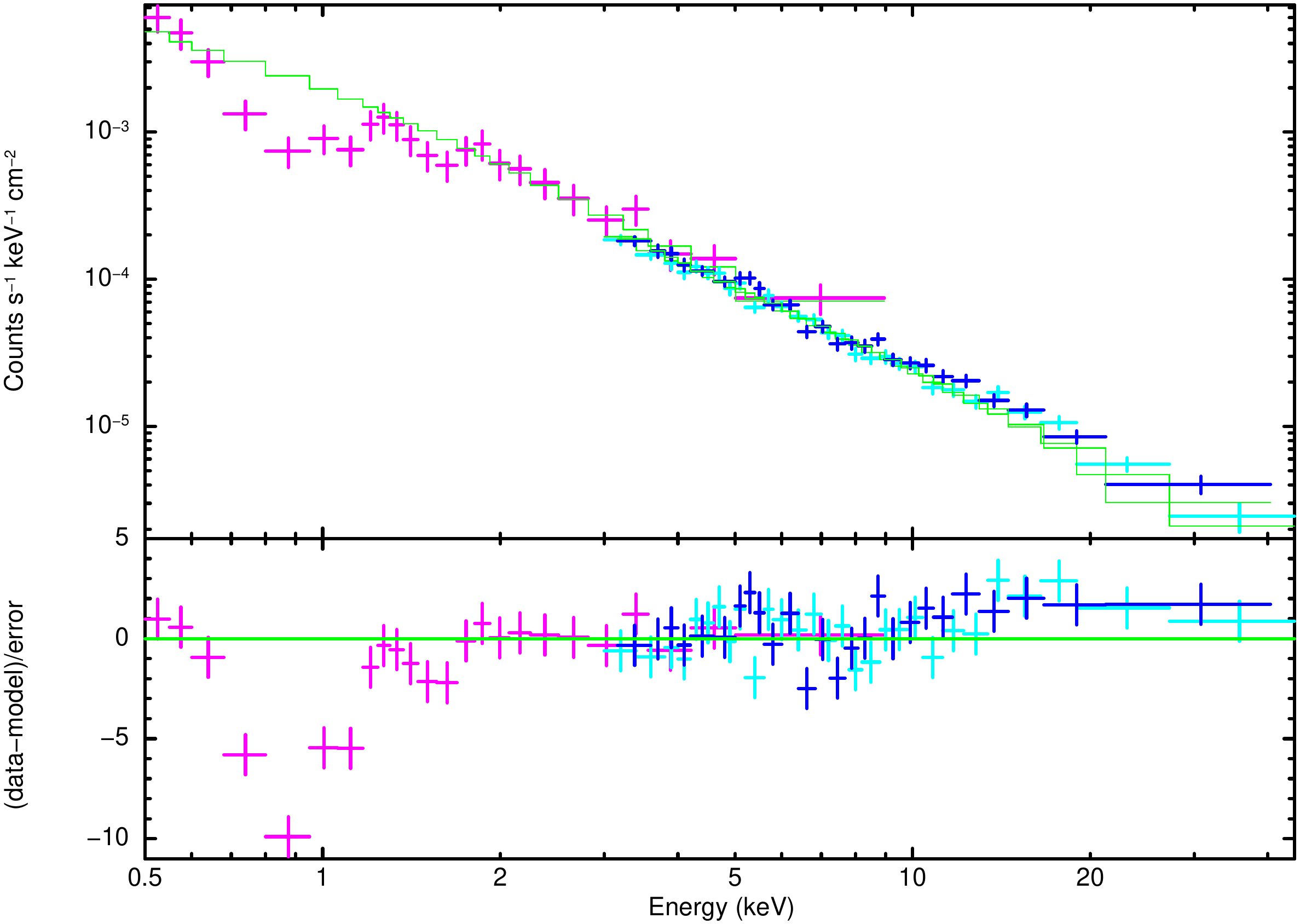}
	\caption{\small{The simultaneous \textit{Swift-NuSTAR} observation of 1E 0754.6+392.8 as fitted by a simple power-law in the 3-10 keV band. Residuals are displayed in both soft and hard X-rays. \label{nupo}}}
\end{figure}

\begin{figure}
	\includegraphics[width=0.48\textwidth]{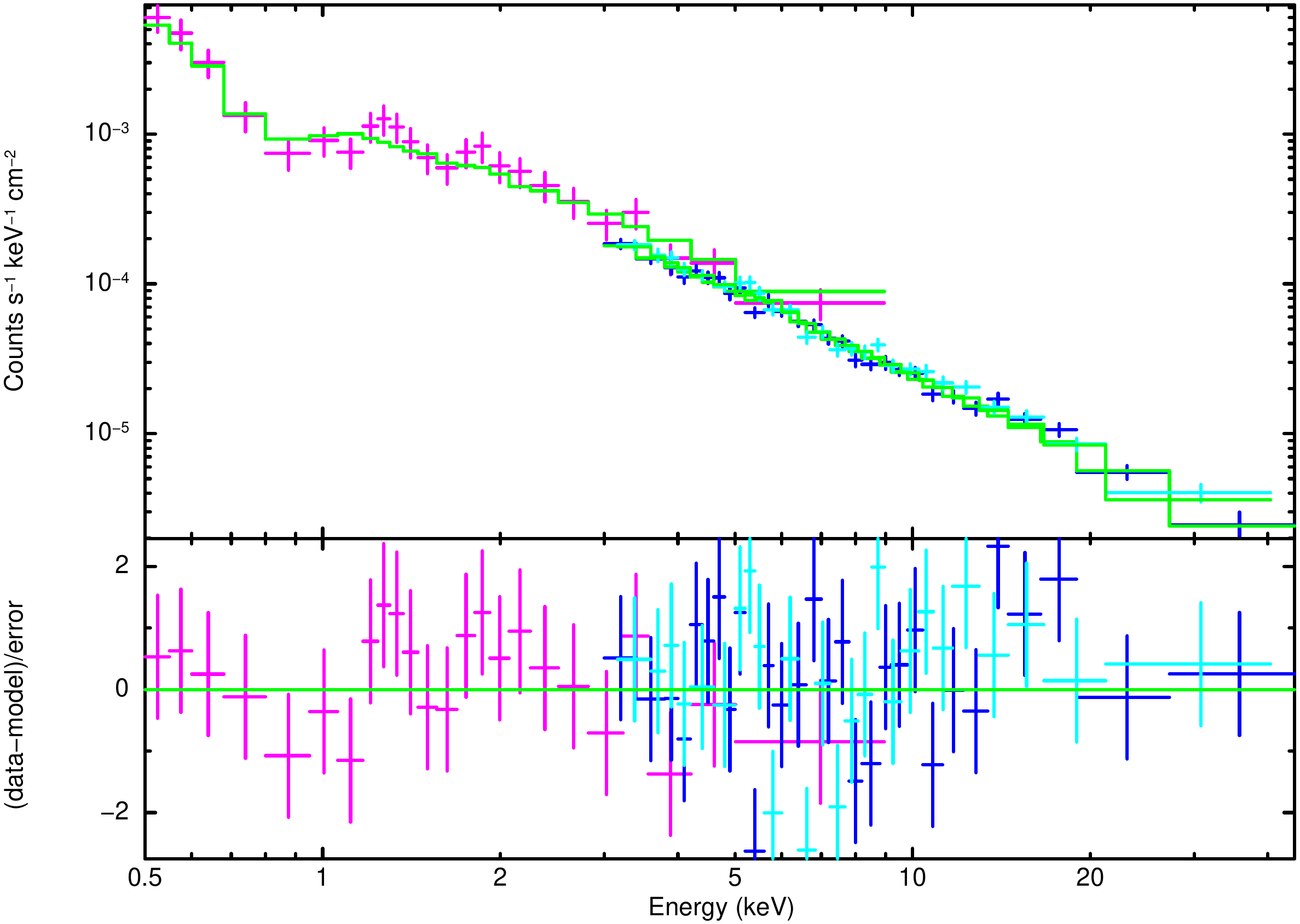}
	\caption{\small{The broadband best-fit ($\chi^2$=174 for 172 d.o.f.~) to the \textit{Swift-NuSTAR} observations is shown.\label{nuxi}}}
\end{figure}

To model the primary continuum, we simultaneously fit the \textit{Swift/NuSTAR} data with a power law in the 3-10 keV band. Moreover, a cross-normalisation constant is included as a free parameter to account for the different instruments involved. This crude model leads to a good fit characterised by $\chi^2$=71 for 66 d.o.f. and a corresponding $\Gamma$=1.95$\pm$0.04 and normalisation N$_{\rm{po}}$=(2.5$\pm$0.2)$\times$10$^{-3}$ photons kev$^{-1}$ cm$^{-2}$ s$^{-1}$. The cross-calibration constants for \textit{Swift} and \textit{NuSTAR} are found to be consistent within $\sim$10\%, while the two \textit{NuSTAR} modules agree each other within a $3\%$.\\
\indent When considering the 0.5-79 keV data, the tested model turns out to be unacceptable in terms of statistics ($\chi^2$=247 for 176 d.o.f.~), mainly due to absorption affecting the soft X-rays, see Fig.~\ref{nupo}. To model this absorption, we included a detailed grid computed with the photoionisation code \textit{XSTAR} \cite[][]{Kall01}. This table takes into account absorption lines and edges for all the metals characterised by an atomic number $Z\leq$30. The \textit{XSTAR} table is calculated assuming a typical $\Gamma$=2 for describing the spectral energy distribution in the 0.1-10$^{6}$ eV band, a high energy cut-off at E$_{\rm{c}}$=100 keV, and a covering factor of 1. Elements abundance is set to the Solar one \cite[][]{Alsplund09}, and a turbulence velocity v$_{\rm{t}}\simeq$ 200 km s$^{-1}$ is considered, based on the typical values of turbulent velocity for warm absorbers \citep{Laha2014}. By letting the ionisation parameter and the column density free to vary, keeping the redshift fixed, the fit is improved by a $\Delta\chi^2$=55 for 2 d.o.f. less. The photon index of this new model is consistent within the errors with what previously obtained. The best-fit values of the WA are $\log(\xi/{\rm erg~s^{-1}~cm})$=2.1$\pm$0.2 and N$_{\rm{H}}$=(3$\pm$1)$\times$ 10$^{22}$ cm$^{-2}$.\\
\indent We further tested the current dataset adopting \textit{xillver} \citep{Garc14a,Daus16}, a self-consistent model reproducing the continuum and ionised reflection of AGNs. In the fit, the photon index, the high energy cut-off, the reflection fraction and the normalisation are left free to vary. The iron abundance A$_{\rm Fe}$ is fixed to the Solar value, while the ionisation parameter $\xi$ is set to the lowest value allowed by the model, close to neutrality. These steps lead to the best-fit ($\chi^2=$ 174 for 172 d.o.f.) showed in Fig. ~\ref{nuxi}. The photon index and the reflection parameter are $\Gamma=2.07\pm0.05$ and $R=0.5\pm0.2$, respectively. A lower limit of E$_{\rm{c}}$ > 170 keV is found for the high energy cut-off while the normalisation is N$_{\rm{xill}}$=(5.0$\pm$0.5)$\times10^{-5}$ photons kev$^{-1}$ cm$^{-2}$ s$^{-1}$. Finally, we find that the WA parameters are still consistent within the errors with the previous fit.

\subsection{\textit{XMM-Newton}: The 0.3-10 keV band, analysis using a phenomenological model}
\begin{figure*}	
	\includegraphics[width=0.98\textwidth]{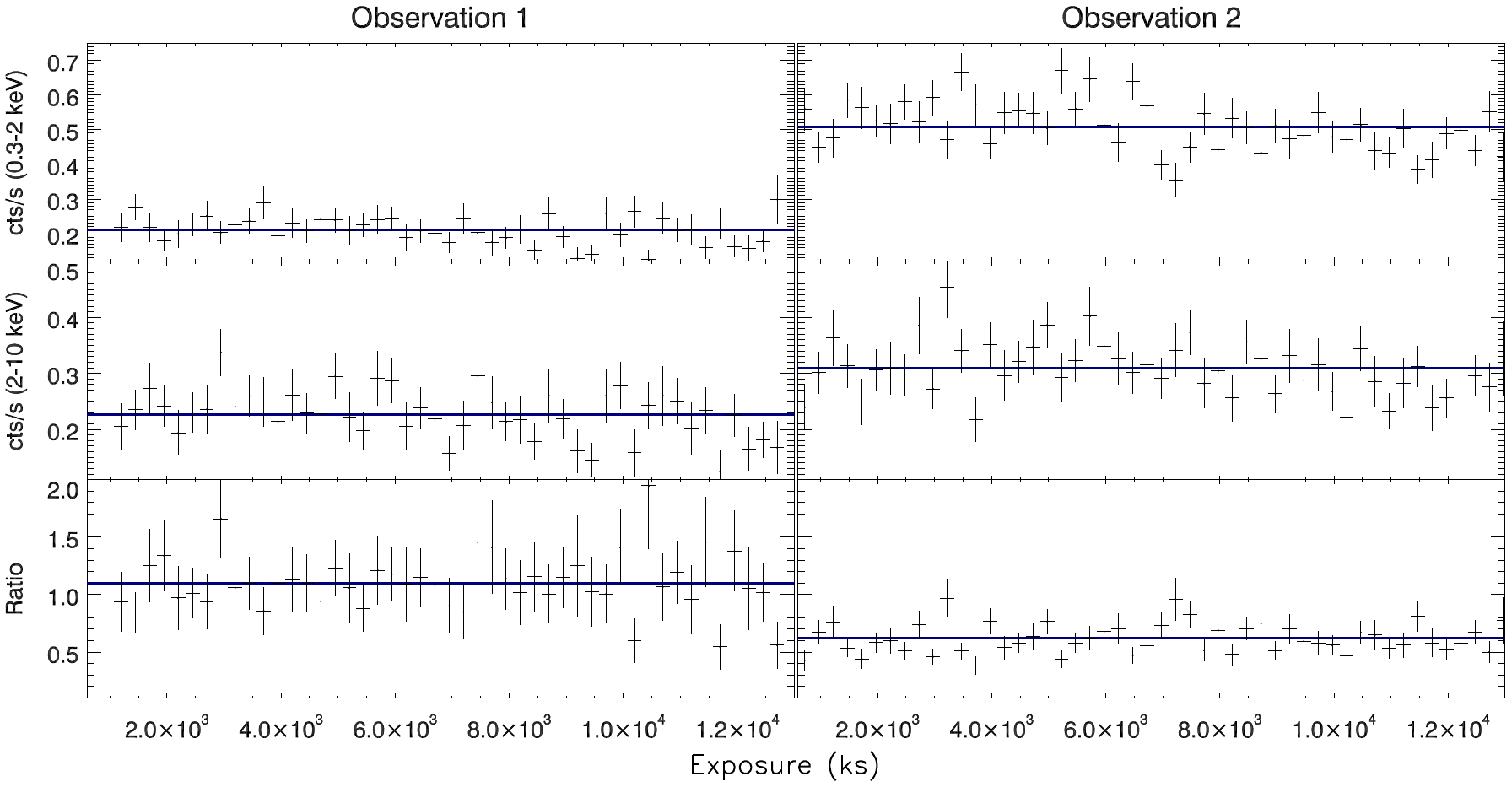}
	\caption{\small{The soft (0.3-2 keV) and hard (2-10 keV) background subtracted \textit{XMM-Newton} light curves and their ratio are displayed. The straight blue lines account for the average rates. Left side panel refers to Obs. 1 while right hand graphs accounts for Obs. 2. The adopted bin is 250 seconds and panels share the same scale. A very weak intra-observation variability, is accompanied by remarkable flux variations between the pointings. In a similar fashion, hardness ratio are constant on ks timescales, while the source shows two different spectral states between Obs. 1 and Obs. 2.}\label{lc}}
\end{figure*}
Visual inspection of the \textit{XMM-Newton} light curves ( see Fig.~\ref{lc}) shows no evidence of flux and spectral variability, hence we used the averaged spectra to improve spectral statistics. On the other hand, the comparison of the hardness ratios between the \textit{XMM-Newton} Obs. 1 and Obs. 2 suggests that the source changed its spectral shape between the two \textit{XMM-Newton} visits.

\begin{figure}
    \includegraphics[width=.49\textwidth]{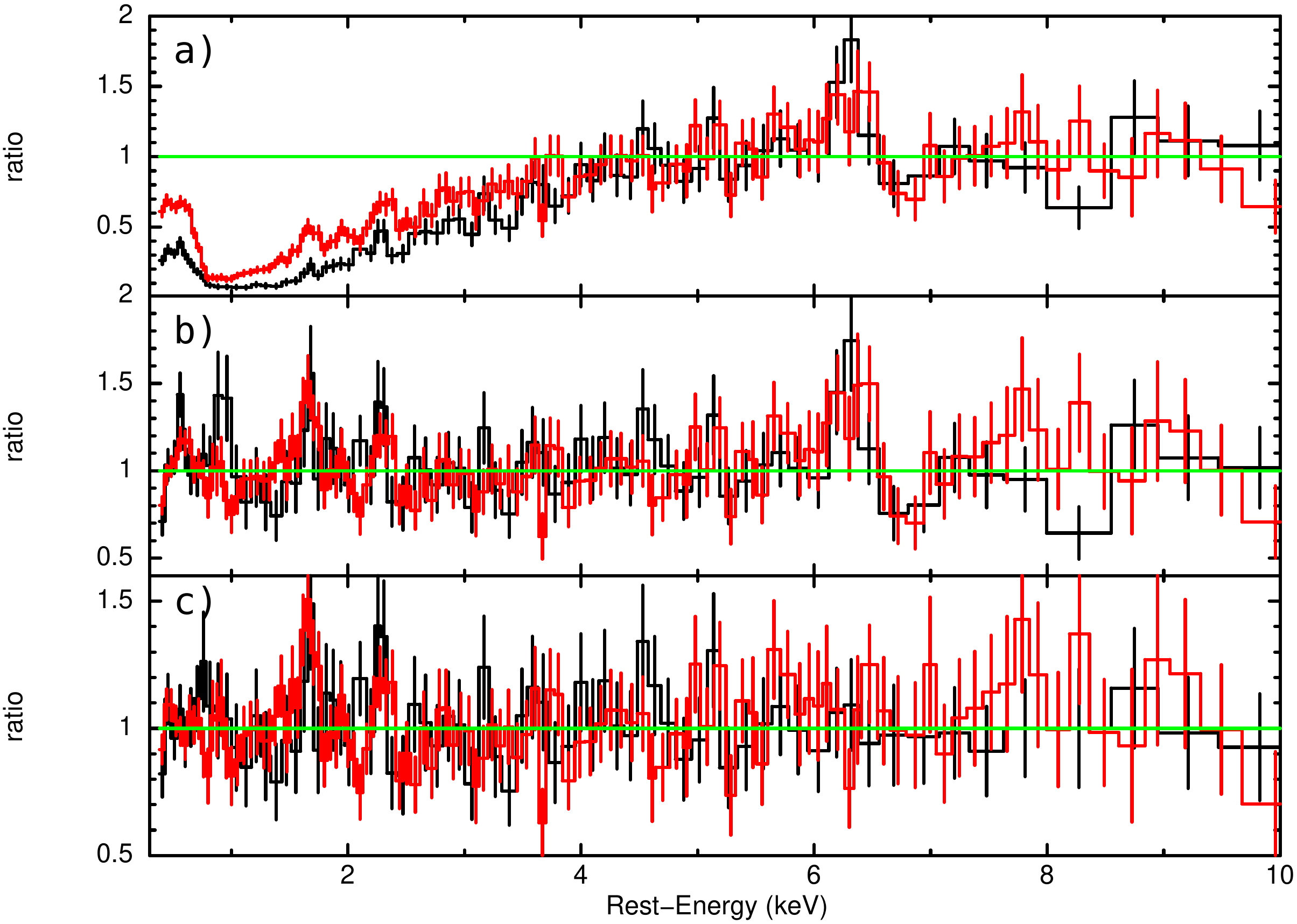}
	\caption{\small{Panel a: The \textit{XMM-Newton} spectra with respect to a power-law only modelling in the 4-10 keV band. Panel b: Low and high energy residuals still present after the WA was included in the modelling. Panel c: Final data-to-model ratio after the inclusion of Gaussian components. A few unmodelled features still populate the energy range between 1.5 and 2.5 keV. However some of these may be directly attributed to the detector systematics. \label{res_model}}}
\end{figure}

\begin{figure}
	\includegraphics[width=.49\textwidth]{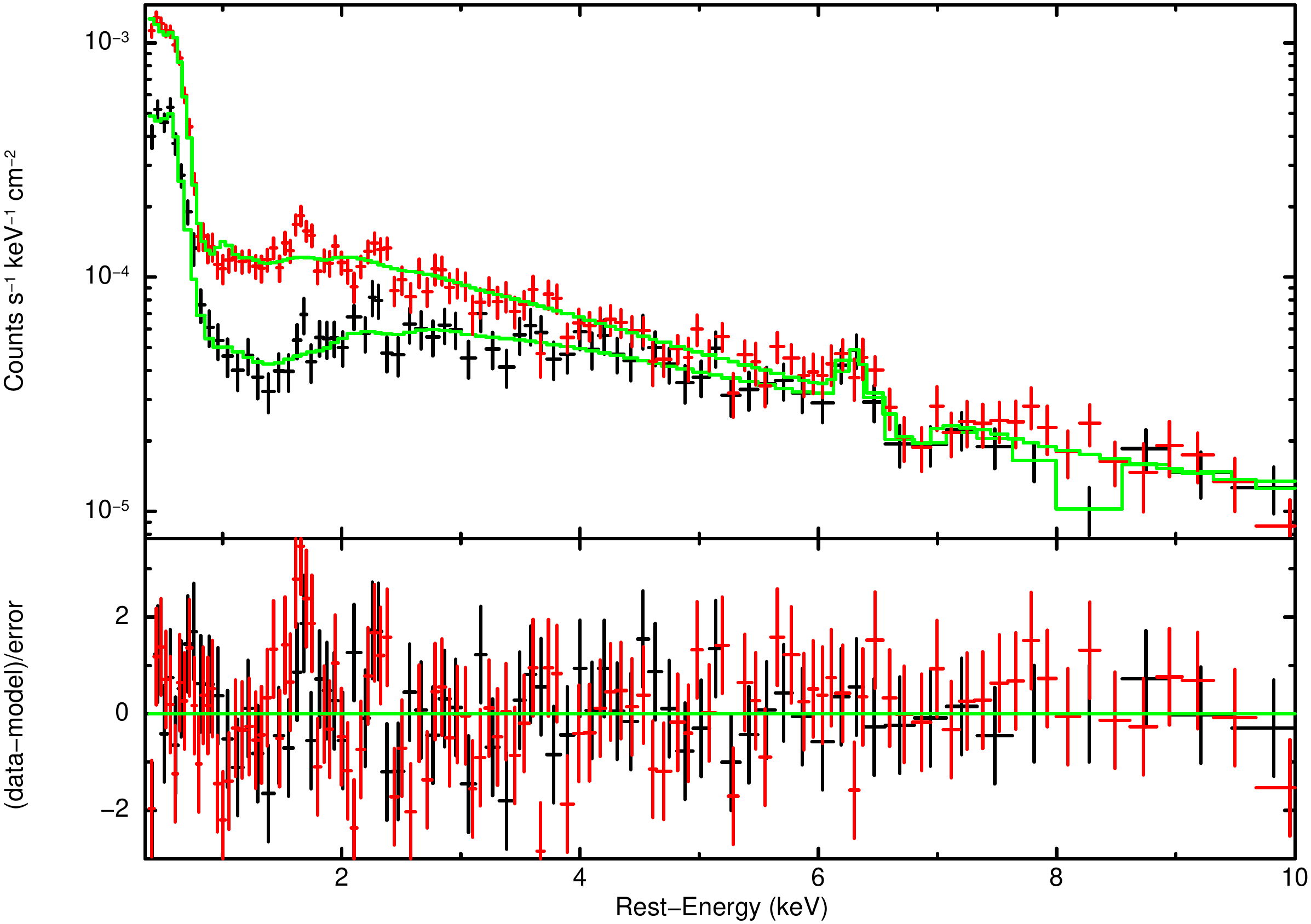}
	\caption{\small{The \textit{XMM-Newton} \textit{EPIC-pn} spectra best-fitted by the phenomenological model including an absorbed power-law and 4 Gaussian components. The corresponding residuals are shown.}} \label{pheno_model}
\end{figure}

\indent The soft X-rays in the \textit{XMM-Newton} observations show signs of intense absorption (see Fig.~\ref{unf} and Fig.~\ref{res_model}, panel a), therefore, we modelled the source spectra using a power-law to reproduce the underlying nuclear continuum and a \textit{XSTAR} table to model the absorption below $\sim$2-3 keV.
In the fit, the photon index the normalisation as well as the WA column density and ionisation are free to vary and untied between Obs. 1 and Obs. 2. The \textit{XSTAR} table enhances the fit statistics by a $\Delta \chi^2/$d.o.f=2258/4 and the resulting modelling (still unacceptable in terms of statistics, $\chi^2$=283 for 179 d.o.f.) is reported in Fig.~\ref{res_model}, panel b. A prominent emission line (which its rest frame energy likely corresponds to the \ion{O}{VII} transition) remains unmodelled in both pointings. We included a Gaussian line to account for it, and by fitting its energy centroid and normalisation (untied between the pointings) a corresponding $\Delta \chi^2$=49 for 4 parameters is found. The current model provides a fit characterised by $\chi^2$=234 for 175 degree of freedom.\\ \indent The present model allows us to focus on the iron line energy band and, in particular, on the absorption/emission features left unmodelled in the 6-9 keV energy interval (see Fig.~\ref{res_model}, b panel). Besides the residuals at about $\sim$6.4 keV in emission, an absorption trough appears in the \textit{XMM-Newton} data at the energy of $\sim$7 keV. Moreover, at higher energies, the first  \textit{XMM-Newton} observation shows a further absorption feature at about $\sim$8 keV. We account for these additional spectral complexities including Gaussian lines: one is used to model the Fe k$\alpha$ emission line and an other to reproduce the absorption feature at E$\simeq$6.8 keV. During the fitting procedure, we assumed that the width of both Gaussian components is zero (unresolved by the instrument resolution) and we let the energy centroid and normalisation free to vary. We find the values for both the emission and absorption lines to be consistent between the two observations. For this reason, we fit these two Gaussian components tying the energy centroid and the normalisation between the two \textit{XMM-Newton} exposures.
From a statistical point of view, the inclusion of the emission line accounting for the Fe K$\alpha$ enhances the fit by $\Delta\chi^2$= 18 for 2 d.o.f., while the absorption line at 6.8 keV leads to a fit improvement of $\Delta\chi^2$= 11 for 2 d.o.f.. These steps yield a global statistics of $\chi^2$=205 for 171 degrees of freedom.
Moreover, as shown by Fig.~\ref{res_model} panel b, a drop of counts is observed at about 8 keV in Obs.~1. We include an additional absorption line in our model to account for it. This line has a free energy centroid and normalisation while its intrinsic width is set fixed to 200 eV, which is comparable to the energy resolution of the \textit{EPIC-pn} at these energies. A $\Delta \chi^2$=8 for 2 d.o.f. indicates that this component is marginally detected.\\
\indent Best-fit parameters for the primary continuum and the Gaussian emission/absorption lines are reported in Table \ref{tabpheno}.
Significant spectral variability is found between Obs. 1 and 2, and the power-law normalisation is found about to double in the second pointing. The observed 2-10 keV fluxes are $(2.0\pm0.1)\times10^{-12}$ and (2.3$\pm0.1)\times10^{-12}$ ergs cm$^{-2}$ s$^{-1}$, respectively, and in the soft X-rays (0.3-2 keV) we find (3.0$\pm0.2)\times10^{-13}$ and (7.4$\pm0.4)\times10^{-13}$ ergs cm$^{-2}$ s$^{-1}$.
Within the errors, the intensity of emission line at energy $\sim$0.57 keV remains constant, and the neutral Fe K$\alpha$ is consistent with being narrow. The Fe K$\alpha$ equivalent width (EW) is also constant between the observations with an average value of $\sim$130 eV. The Gaussian line at $\sim$6.8 keV is likely associated with blueshifted \ion{Fe}{XXV}, while the marginally detected component at higher energy is more likely associated with a highly blueshifted \ion{Fe}{XXVI}.\\
\indent In panel c of Fig.\ref{res_model} and in Fig.\ref{pheno_model} we notice the presence of further features not reproduced by the current phenomenological model, especially between 1.5-2.5 keV. Some of these features in this band may be directly attributed to the detector calibration uncertainties (e.g. Si K-edge 1.84 keV), and at the Au M-edge ($\sim$2.4 keV) \citep[see][for discussions and comparisons]{Kaas11,DiGe15,Ursi15,Capp16,Middei18}. We notice that modelling these features with Gaussian lines or ignoring the spectra in the 1.5-2.5 keV band do not affect the values reported in Table 2.

 \begin{table}
 	\setlength{\tabcolsep}{4.5pt}
 	\caption{\small{Parameters for the best-fit model including fully covering ionised absorption and Gaussian lines. The power law normalisation (normalised at 1 keV) is in units of photons keV$^{-1}$ cm$^{-2}$ s$^{-1}$, while units for the column density and ionisation parameter are 10$^{22}$ cm$^{-2}$ and erg s$^{-1}$ cm, respectively. The normalisation of emission and absorption lines are in photons cm$^{-2}$ s$^{-1}$. Finally, a $\dagger$ is used to identify the parameters kept frozen during the fit while the $\star$ specifies the component whose parameters were tied between the observations.}\label{tabpheno}}
 		\label{fitnoline}
 	\begin{tabular}{c c c c}
 		\hline
 		\setlength{\tabcolsep}{1.pt}
 		Component & Parameter & Obs. 1& Obs. 2 \\
 		\\
 		\hline
 		\hline	
 		TBabs & N$_{\rm{H}}\dag$ & 0.056 & 0.056 \\
 		\hline
 		power-law & $\Gamma$ &1.66$\pm$0.04 &2.05$\pm$0.10 \\
 		& Norm ($\times$10$^{-3}$)   &0.64$\pm$0.1 &1.4$\pm$0.2\\
 		\hline
 		WA      	&$\log\xi$     &2.00$\pm$0.04&1.75$\pm$0.15 \\
 		&$N_{\rm{H}}$   &7.9$\pm$0.9 &5.4$\pm$0.4   \\
 		&$z\dag$         &0.096&  \\
 		&$\Delta\chi^2$/d.o.f.    &642/2&1616/2\\
 		\hline
 		zgauss  (Emi)  &E (keV)  &0.56$\pm$0.02& 0.59$\pm$0.02 \\
 		&$z\dag$     &0.096& -  \\
 		&Norm      ($\times10^{-4}$)   &1.4$\pm$0.4&2.6$\pm$1.0 \\
 		&EW (eV)    &70$\pm$30&  45$\pm$25 \\
 		&$\Delta\chi^2$/d.o.f.    &18/2&31/2\\
 		\hline
 		zgauss (Emi)$\star$   &E (keV) & 6.30$\pm$0.8  & \\
 		&$z\dag$    &0.096&   \\
 		&Norm   ($\times$10$^{-6}$)  &5.0$\pm$2.0&   \\
 		&EW (eV)    &130$\pm$90&125$\pm$80  \\
 		&$\Delta\chi^2$/d.o.f.    &18/2&\\
 		\hline
 		zgauss  (Abs)$\star$  &E (keV)    &6.80$\pm$0.07 &  \\
 		&$z\dag$      &0.096&   \\
 		&Norm ($\times$10$^{-6}$)         &-3.5$\pm$1.2&  \\
 		&EW (eV)    &-100$\pm$50&-100$\pm$50   \\
 		&$\Delta\chi^2$/d.o.f.    &11/2&\\
 		\hline
 		zgauss  (Abs)  &E (keV)  &8.2$\pm$0.2&   \\
 		&$\sigma\dag$  (eV)   &200& -  \\
 		&$z\dag$     &0.096& -  \\
 		&Norm      ($\times10^{-6}$)   &-5.4$\pm$3.1&  - \\
 		&EW (eV)    &-220$\pm$120&  - \\
 		&$\Delta\chi^2$/d.o.f.    &8/2&\\
 	\end{tabular}
 \end{table}

\subsection{\textit{XMM-Newton}: The 0.3-10 keV band, analysis using a physical model}

\begin{figure}
	\includegraphics[width=0.49\textwidth]{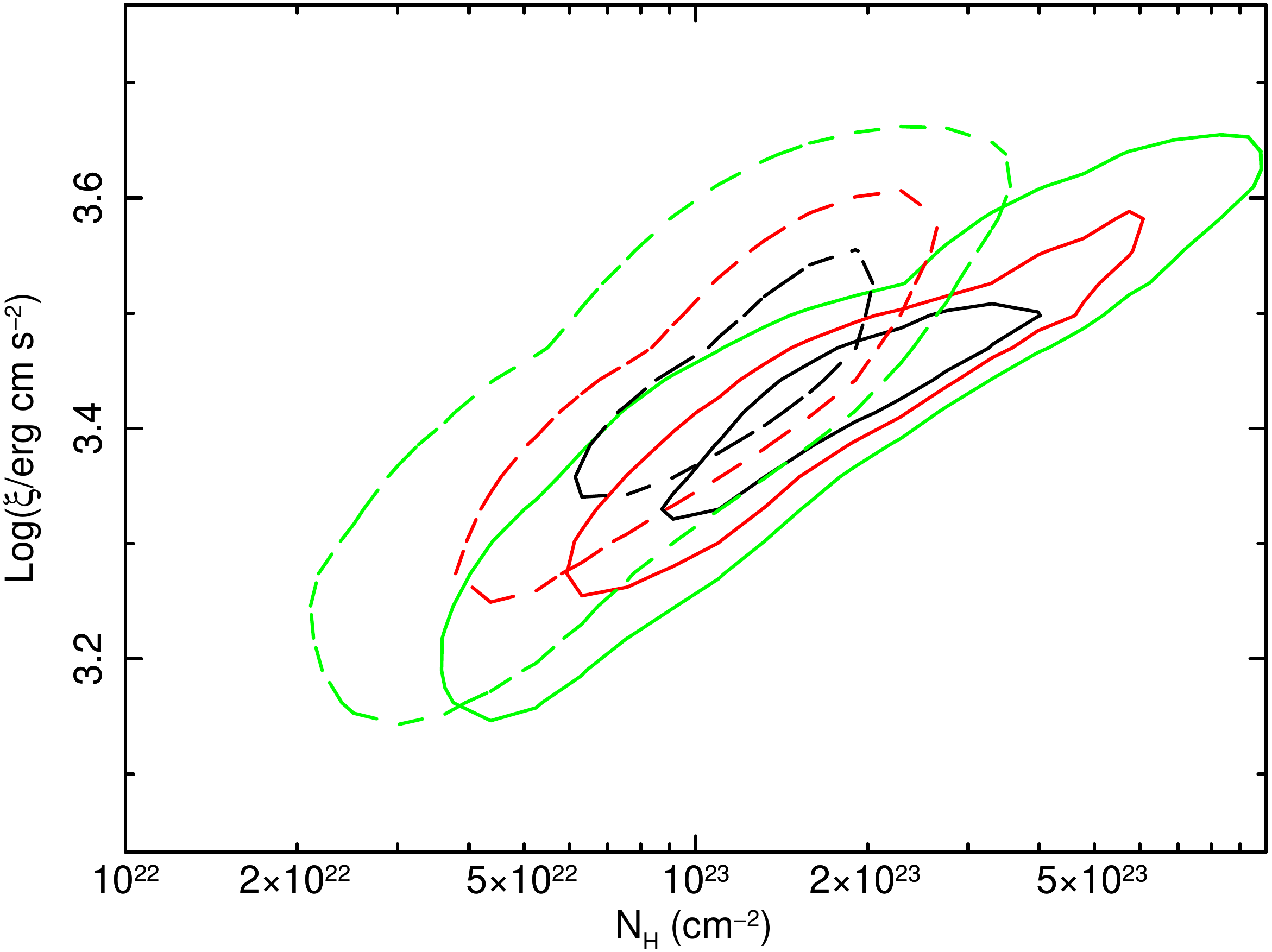}
	\caption{\small{The contours at 99\% (green), 90\%(red) and 68\% (black) confidence level computed for Abs1 (solid lines) and Abs2 (dashed lines).}\label{abs12}}
\end{figure}

\begin{figure}
	\includegraphics[width=0.49\textwidth]{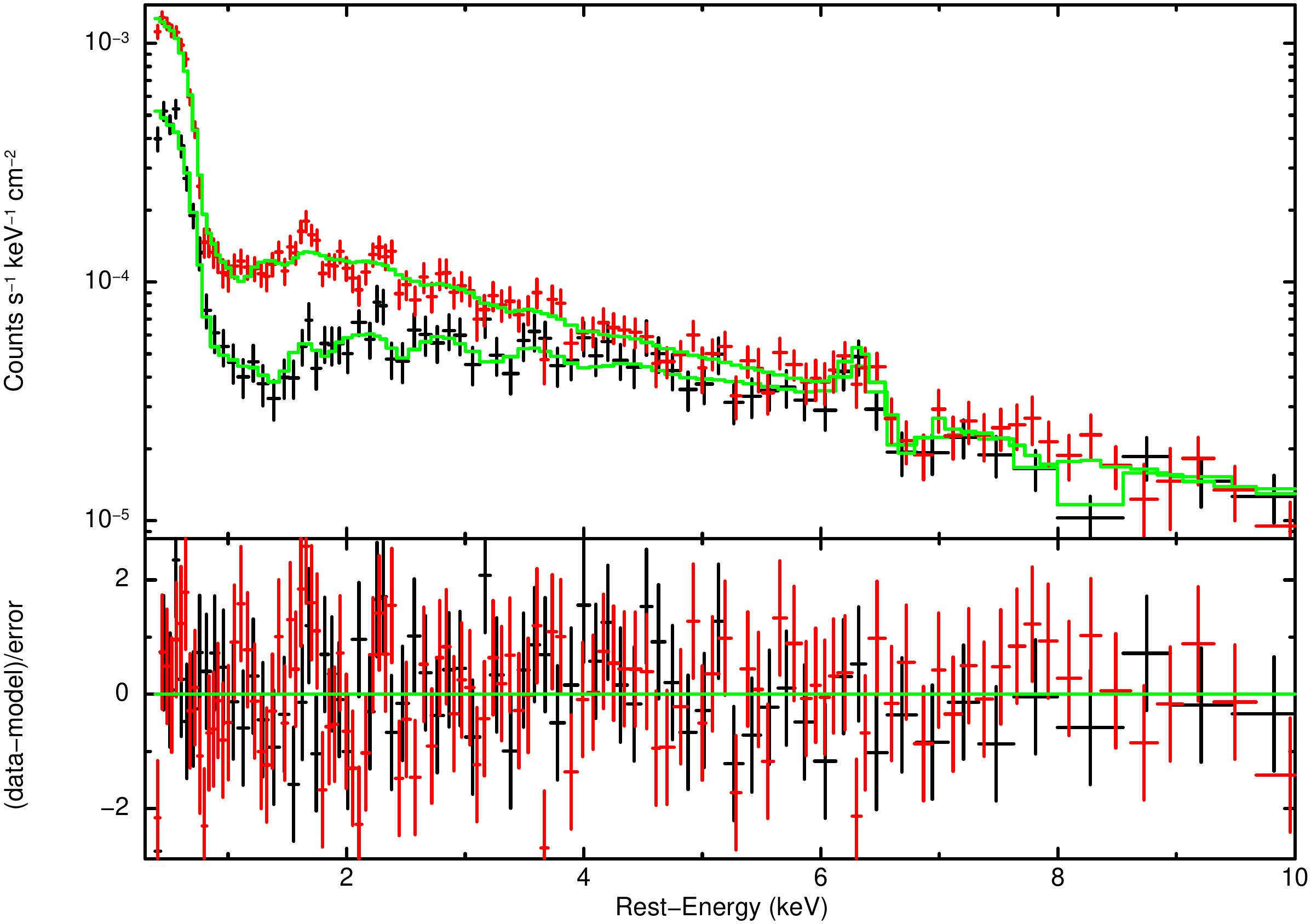}
	\caption{\small{The best-fit model to the two \textit{XMM-Newton} observations of the model including three different \textit{XSTAR  } tables and \textit{xillver}.}\label{best_xillver}}
\end{figure}

As a subsequent step we reanalysed the \textit{EPIC-pn} spectra using a self-consistent emission model (\textit{xillver}). Such a model simultaneously fits the source emission and its associated ionised reflection component. Moreover, we accounted for the absorption troughs in the 6.5-8.5 keV energy band including two \textit{XSTAR} tables. One, Abs1, is used to reproduce the absorption at E$\sim$6.8 keV, while, the other, Abs2 is included to model the  absorption at 8 keV.
The fit was performed allowing the photon index, the reflection fraction and the normalisation to vary also between the two pointings. The high energy cut-off is kept frozen to $E_{\rm{c}}$=100 keV, while the ionisation parameter is free to vary but tied between Obs. 1 and Obs. 2. Concerning the ionised absorbers, we fit the ionisation parameter and column density in both the observations. The table accounting for the drop at 6.8 keV was fitted tying its parameters between the two \textit{XMM-Newton} observations, while the grid accounting for Abs2 was only included in Obs. 1. Finally, we considered the redshift of all the tables (z$_{\rm{obs}}$) as being a free parameter in order to constrain the possible velocity shift.\\ \indent The steps just described lead to the best-fit in Fig.~\ref{best_xillver}. The fit has a statistics of $\Delta\chi^2$=170 for 168 d.o.f.~ and the corresponding best-fit values of the various parameters are reported in Table ~\ref{xstartab}. As already shown by the phenomenological model, spectral variability characterises the primary continuum of 1E 0754.6+392.8. The nuclear emission normalisation increases by a factor $\sim$1.25 between the two \textit{XMM-Newton} visits, while the reflection fraction $R$ is found to be $\sim$1 and constant within the uncertainties. The WA component varies both in column density and ionisation in the range 3$\times10^{22} \rm cm^{-2}\lesssim$ N$_{\rm{H}}\lesssim7\times10^{22} \rm cm^{-2}$ and  1.5$\lesssim\log(\xi/{\rm erg~s^{-1}~cm})\lesssim$2.1, respectively. These values, though slightly smaller, are consistent within the errors with those of the phenomenological model. Higher ionisations states and column densities characterise both Abs1 and Abs2. The physical parameters of these two components are marginally constrained by the current dataset and in Fig.~\ref{abs12} we show the confidence regions for the $\log\xi$ vs N$_{\rm H}$ parameters. Using the redshift best-fit values, we can only found an upper limit of v$_{\rm out}\leq$1500 km s$^{-1}$ for the WA.
Abs1 is consistent with being in outflow with a velocity v$_{\rm out}$ in the range 4400-6200 km s$^{-1}$, while a v$_{\rm{out}}$=(0.23$\pm$0.02)$c$ ($c$ being the speed of light) is estimated for the possible Abs2 component.
These physical quantities for various absorbers in 1E 0754.6+392.8 are compatible with what is found by other authors \citep[e.g.][]{Tomb10,Tomb13} with the exception of the WA column density that appears to be larger than typical values, but it could be related to a more equatorial inclination of this AGN \citep[see also][]{Krolik_2001,Behar17}.
\begin{table}
	\centering
	\setlength{\tabcolsep}{2.pt}
	\caption{\small{Values and corresponding uncertainties for the best-fit parameters are shown. The overall fit statistics is $\Delta\chi^2$=170 for 168 d.o.f. . The $\Delta \chi^2$ and the corresponding variation of degrees of freedom are also reported}. For Abs1 and Abs2 the turbulence velocities are 300 km s$^{-1}$ and 10000 km s$^{-1}$, respectively}\label{xstartab}
	\begin{tabular}{c c c c}
		\hline
		
		Model & Parameter& Obs. 1& Obs. 2 \\
		\\
		\hline
		\hline
		TBabs & N$_{\rm{H}}\dag$ & 0.056 & 0.056 \\
		\hline
		xillver 		
		&$\Gamma$&1.66$\pm$0.04&2.07$\pm$0.06\\
		&$\log\xi\dag$&1.3$^{+0.1}_{-0.2}$&\\
		&R&1.0$\pm$0.5&1.0$\pm$0.4\\
		&Norm ($\times$10$^{-5}$)&1.5$\pm$0.2&2.0$\pm$0.1\\
		\hline
		WA & $\log\xi$&2.00$\pm$0.05&1.50$\pm$0.07\\
		&N$_{\rm{H}}$ ($\times$10$^{22}$ cm$^{-2}$)
		&7.0$\pm$1.0&4.0$\pm$0.5\\
		&v$_{out}$/$c$&<0.005&<0.004\\
		&v$_{out}$ (km s$^{-1}$)&<1500&<1100\\
		&$\Delta\chi^2/d.o.f.$&482/3&1171/3\\ 
		\hline
		Abs1 $\star$ & $\log\xi$&3.4$\pm$0.1&\\
		&N$_{\rm{H}}$ ($\times$10$^{23}$ cm$^{-2}$)
		&2.6$^{+2.2}_{-1.8}$&\\
		&v$_{out}$/$c$&0.017$\pm$0.04&\\
		&v$_{out}$ (km s$^{-1}$)&5300$\pm$900&\\
		&$\Delta\chi^2/d.o.f.$&23/3&\\ 
		\hline
		Abs2& $\log\xi$&3.4$\pm$0.3&-\\
		&N$_{\rm{H}}$ ($\times$10$^{23}$ cm$^{-2}$)
		&1.3$^{+1.0}_{-0.8}$&-\\
		&v$_{out}$/$c$&0.23$\pm$0.02&-\\
		&v$_{out}$ (km s$^{-1}$)&74000$\pm$10000&\\
		&$\Delta\chi^2/d.o.f.$&14/3&-\\ 

	\end{tabular}
\end{table} 

\subsection{Comparison between \textit{XMM-Newton} and \textit{Swift/NuSTAR} data}
\begin{figure}	
	\includegraphics[width=0.49\textwidth]{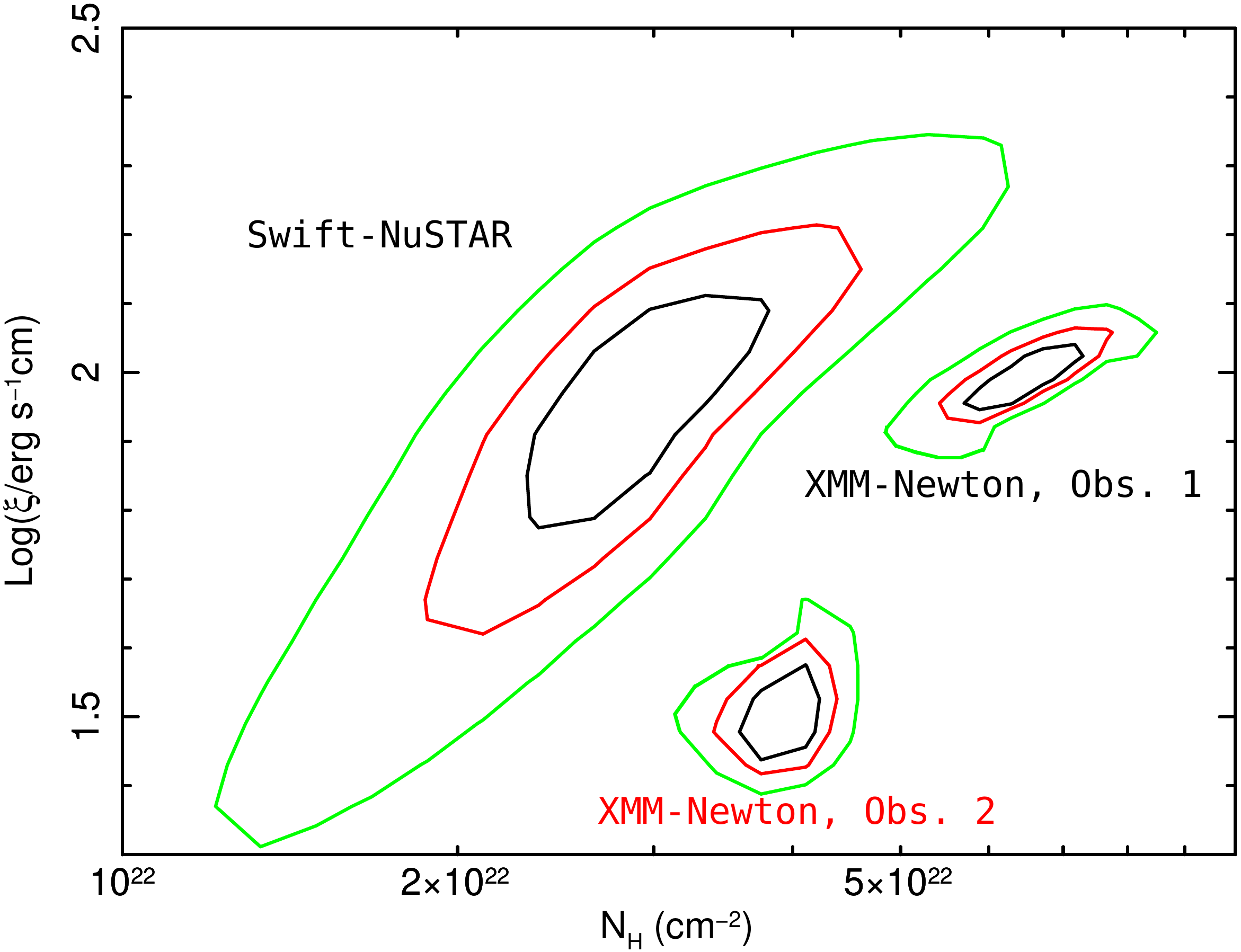}
	\caption{\small{The 68\% (black), 90\% (red) and 99\% (green) confidence level contours are plotted for the column density and ionisation state of the WA component in 1E 0754.6+392.8, respectively. Such a component is clearly monthly variable and, though with higher uncertainties due to the poorer statistics of the \textit{XRT} observation, over the years.}\label{contours}}
\end{figure}

The current dataset covers a $\sim$8 years long time interval, thus it is suitable for variability studies. As showed in Fig.~\ref{unf}, the spectra of 1E 0754.6+392.8 varied both in shape and amplitude. The observed flux in the 0.5-2 keV band exhibits a change between the \textit{XMM-Newton} exposures and in the subsequent \text{Swift-NuSTAR} pointing in which it increased by a factor larger than 10. On the other hand, the observed 3-10 keV flux is fairly consistent with F$_{\rm{3-10~keV}}\sim2\times$10$^{-12}$ erg cm$^{-2}$ s$^{-1}$ in Obs. 1 and Obs. 2, while the source doubled the flux in the same band during the \text{Swift-NuSTAR} observation. It is worth noting that the reflection fraction $R$ is higher when the 3-10 keV flux is lower, suggesting a constant reflected emission.
Intense absorption in the soft X-rays is the major component that shapes the source spectrum.\\ \indent The WA varies in ionisation and column density. For the physical parameters of this component and using the self-consistent models in Sect. 3.1 and 3.3 we computed the confidence regions showed in Fig.~\ref{contours}. These contour plots have been calculated by assuming the redshift of the \textit{XSTAR} tables fixed at its best-fit value. In Fig.~\ref{contours}, the low statistics of the \textit{XRT} data and the \textit{FPMA\&B} bandpass explain the poor constraints on the WA in the 2014 observation.\\
\indent Moreover, we tested \textit{Swift-NuSTAR} data for the presence of absorption lines. We started with the phenomenological model presented in Sect. 3.1 to which we have added Gaussian components in absorption: a narrow line with an energy centroid of 6.8 keV; a 200 eV width absorption line centered at 8.2 keV. The fit to the data does not require a Gaussian component at 6.8 keV. On the other hand, the other line provides a weak improvement of the fit ($\Delta\chi^2$/d.o.f.=6/2) and the absorption line is characterised by E=8.9$\pm$0.3 keV, N=(6.2$\pm$4.0)$\times$10$^{-6}$ photons cm$^{-2}$ s$^{-1}$ EW=-100$\pm$60 eV. However, the statistics of the available \textit{Swift-NuSTAR} exposure is not suitable for adequately searching for faint absorption features.
\section{Statistical significance of the absorption features}

In order to assess the statistical significance of the two absorption features at $\sim$6.8 and $\sim$8 keV, we performed Monte Carlo simulations. These simulations are particularly suitable for quantifying the correct significance of any absorption/emission component detected with a blind search over a certain energy interval \citep[e.g.][]{Porquet2004,Miniutti2006,Markowitz2006,Tomb10,Goff13,Tombesi2014,Marinucci2018,Smith2019}.
Therefore, we used the \textit{fakeit} command in \textit{XSPEC} to generate a set of 1000 synthetic spectra for each of the two exposures. To simulate these fake spectra, we used the background and response files of the real data and the same exposure time of the observations. The underlying model considered for the simulations is the one presented in Sect. 3.2 but without the emission and absorption lines. Finally, the simulated data were binned in the same way as those observed. After that, we add a new narrow (or with a width of 200 eV, for the candidate UFO) Gaussian line whose normalisation was initially set to zero and free to vary in the range between -1 and +1. The energy centroid was free to vary between 6.5 and 9 keV for both the features, in order to sample the searched energy interval. We used the \textit{steppar} command in \textit{XSPEC} to map the $\Delta\chi^2$ using 100 eV steps, and the resulting variation where recorded. \\
\indent Defining N as the number of simulations in which a chance improvement of the $\chi^2$ is found to be equal or larger than the one on the real data and S being the total number of simulated spectra, we estimated the Monte Carlo statistical significance of the detections to be 1-N/S. Following this definition, we obtained N=16 with S=1000 for the absorption line at 6.8 keV. Hence the significance of this feature in accordance with the simulations is 98.4\%, this corresponding to a 2.4$\sigma$ detection. For the candidate UFO at E$\sim$8 keV, we find N=39 which corresponds to a significance of 96.1\% i.e. 2.06$\sigma$. We note that the Monte Carlo statistical significance of these features is higher than the threshold of 95\% typically used in extensive searches of Fe K features \citep[e.g.][]{Tomb10,Tomb14,Goff13}.

\section{Conclusions and summary}

\indent We reported on the first X-ray broadband (0.3-79 keV) spectroscopic analysis of the NLSy1 galaxy 1E 0754.6+392.8 based on two 2006 \textit{XMM-Newton} observations (taken 6 months apart) and on a \textit{NuSTAR}-\textit{Swift} simultaneous snapshot performed in 2014. The spectra of 1E 0754.6+392.8 are well described by a variable power-law spectrum with a photon index between 1.65 and 2.07. This spectral variability is observed from months to years, while down to hours the source exhibits a constant behaviour as suggested by the hardness ratios in Fig.~\ref{Nulc} and Fig.~\ref{lc}). Long term flux variations mainly affect the soft X-rays (0.3-2 keV), whose flux doubles in six months and increases more than a factor 10 in 8 years . On the other hand, the continuum emission at higher energies is less affected by variations on monthly timescales, but, over the years, the observed 2-10 keV flux increased by a factor $\sim$2.5. \\
\indent Two significant emission lines are detected at $\sim$6.4 keV and $\sim$0.57 keV, respectively. The former is interpreted as fluorescent emission of K-shell iron in a low-ionisation state. Its width is unresolved in both the observations (upper limits $\sigma_{\rm{\ion{Fe} K\alpha}}$<0.19 keV), and this may rule out an origin in the inner parts of the accretion disc. Furthermore, such emission feature has a constant equivalent width and normalisation. The reflected flux is consistent with being constant over the years and it is likely originating in distant material. The other feature in emission observed at about $\sim$0.57 keV may result from \ion{He}-like oxygen triplet emission arising from the same low-ionisation state plasma responsible for the iron K emission line. However, such line may also be produced in a much farther region from the SMBH (e.g. the Narrow Line Region) and it is more easily detected due to the presence of the WA itself \citep[][]{Pico04}.\\
\indent The main spectral feature in the spectrum is an intense absorption affecting the soft X-ray band. In particular, data show a variable WA that is persistent over the years. This component is consistent with being at the same redshift of the source and at pc scales with respect to the central SMBH. The physical quantities derived for the WA are shown in Fig.~\ref{contours} for the different observations. The change in column density and/or ionisation state of this component can be the result of a clumpy or filamentary inhomogeneous absorber \citep[e.g.][]{Gaspari2017,Seraf19}. \\
 \indent The iron line energy band shows further signatures of absorption likely due to highly ionised and high column density matter crossing our line of sight. Though, these components have only a low significance ($\sim$98\% for \ion{Fe}{XXV} and $\sim$96\% for \ion{Fe}{XXVI} assessed using Monte-Carlo simulations). The Fe K absorber Abs1 is observed in both XMM-Newton observations and is consistent with a \ion{Fe}{XXV}.	
Abs2, a candidate UFO, is characterised by a mildly outflowing velocity (v$_{\rm{out}}$=(0.23$\pm$0.03)$c$) and its ionisation and column density are compatible with what often observed for UFOs \citep{Tomb11,Parker17,Reeves2018,Parker18,Braito18,Seraf19,Matzeu19}.
\\ \indent The presence/absence of the Abs2 component in the analysed data is consistent with SMBH winds being variable as it has been repeatedly confirmed through ensemble studies or single object analyses. For instance, \cite{Tomb10} reported on the variability of such winds using a sample of Seyfert galaxies and found a detection rate of the order of 50$\pm$20\% for these components \citep[see also][]{Tomb11,Goff13}.\\
\indent It is worth noticing that the various types of absorbers can be part of a single large-scale multiphase outflow seen at different distances from the SMBH \citep[e.g.][]{Tomb13}. The properties of the WAs, the UFOs and the highly ionised non-UFO absorbers (like as our Abs1) have been found to show significant trends: the closer the absorber is to the central BH, the higher the ionisation, column, outflow velocity. Within this context, the possible simultaneous presence of three different types of absorbers suggests 1E 0754.6+392.8 to be a fantastic laboratory in which to study the relations between the different absorbing phases. In fact, though being rarely observed so far, the presence of multiple phases allows unprecedented insights on the outflows structure and physics \citep[e.g.][]{Seraf19}.\\
\indent In conclusion, the current data, despite the low S/N, suggest that 1E 0754.6+392.8, clearly hosting a variable WA, may have further absorbing phases characterised by much higher outflow velocities. Only longer exposures or the higher sensitivity of an X-ray calorimeter (e.g. XRISM and Athena) will allow us to put firmer conclusions on the putative multiphase outflows possibly present on this source and to better assess for the presence of its accretion disc wind component.

\begin{acknowledgements}
We thank the referee for her/his suggestions which improved the manuscript. RM thanks Valentina Braito and James Reeves for useful discussions and comments. RM acknowledges support from the Faculty of the European Space Astronomy Centre (ESAC), Fondazione Angelo Della Riccia for financial support and Universit\'e Grenoble Alpes and the high energy SHERPAS group for welcoming him at IPAG. FT acknowledges support by the Programma per Giovani Ricercatori - anno 2014 “Rita Levi Montalcini”. Part of this work is based on archival data, software or online services provided by the Space Science Data Center - ASI. SB and AZ acknowledge financial support from ASI under grants ASI-INAF I/037/12/0 and n. 2017-14-H.O. RS acknowledges financial contribution from the agreement ASI-INAF n.2017-14-H.0. POP thanks financial support by the french CNES agency. GAM is supported by European Space Agency (ESA) Research Fellowships. This  work  is  based  on  observations obtained with: the NuSTAR mission,  a  project  led  by  the  California  Institute  of  Technology,  managed  by  the  Jet  Propulsion  Laboratory  and  funded  by  NASA; XMM-Newton,  an  ESA  science  mission  with  instruments  and  contributions  directly funded  by  ESA  Member  States  and  the  USA  (NASA).
\end{acknowledgements}

\thispagestyle{empty}
\bibliographystyle{aa}
\bibliography{1E0754_accepted.bib}

\begin{thebibliography}{69}
\expandafter\ifx\csname natexlab\endcsname\relax\def\natexlab#1{#1}\fi

\bibitem[{{Asplund} {et~al.}(2009){Asplund}, {Grevesse}, {Sauval}, \&
  {Scott}}]{Alsplund09}
{Asplund}, M., {Grevesse}, N., {Sauval}, A.~J., \& {Scott}, P. 2009, \araa, 47,
  481

\bibitem[{{Behar} {et~al.}(2017){Behar}, {Peretz}, {Kriss}, {Kaastra}, {Arav},
  {Bianchi}, {Brand uardi-Raymont}, {Cappi}, {Costantini}, \& {De
  Marco}}]{Behar17}
{Behar}, E., {Peretz}, U., {Kriss}, G.~A., {et~al.} 2017, \aap, 601, A17

\bibitem[{{Berton} {et~al.}(2015){Berton}, {Foschini}, {Ciroi}, {Cracco}, {La
  Mura}, {Lister}, {Mathur}, {Peterson}, {Richards}, \& {Rafanelli}}]{Bert15}
{Berton}, M., {Foschini}, L., {Ciroi}, S., {et~al.} 2015, \aap, 578, A28

\bibitem[{{Bianchi} {et~al.}(2009){Bianchi}, {Guainazzi}, {Matt}, {Fonseca
  Bonilla}, \& {Ponti}}]{Bian09}
{Bianchi}, S., {Guainazzi}, M., {Matt}, G., {Fonseca Bonilla}, N., \& {Ponti},
  G. 2009, \aap, 495, 421

\bibitem[{{Blustin} {et~al.}(2005){Blustin}, {Page}, {Fuerst},
  {Branduardi-Raymont}, \& {Ashton}}]{Blustin05}
{Blustin}, A.~J., {Page}, M.~J., {Fuerst}, S.~V., {Branduardi-Raymont}, G., \&
  {Ashton}, C.~E. 2005, \aap, 431, 111

\bibitem[{{Braito} {et~al.}(2018){Braito}, {Reeves}, {Matzeu}, {Severgnini},
  {Ballo}, {Caccianiga}, {Campana}, {Cicone}, {Della Ceca}, \&
  {Turner}}]{Braito18}
{Braito}, V., {Reeves}, J.~N., {Matzeu}, G.~A., {et~al.} 2018, \mnras, 479,
  3592

\bibitem[{{Cappi} {et~al.}(2016){Cappi}, {De Marco}, {Ponti}, {Ursini},
  {Petrucci}, {Bianchi}, {Kaastra}, {Kriss}, {Mehdipour}, {Whewell}, {Arav},
  {Behar}, {Boissay}, {Branduardi-Raymont}, {Costantini}, {Ebrero}, {Di Gesu},
  {Harrison}, {Kaspi}, {Matt}, {Paltani}, {Peterson}, {Steenbrugge}, \&
  {Walton}}]{Capp16}
{Cappi}, M., {De Marco}, B., {Ponti}, G., {et~al.} 2016, \aap, 592, A27

\bibitem[{{Dauser} {et~al.}(2016){Dauser}, {Garc{\'{\i}}a}, {Walton},
  {Eikmann}, {Kallman}, {McClintock}, \& {Wilms}}]{Daus16}
{Dauser}, T., {Garc{\'{\i}}a}, J., {Walton}, D.~J., {et~al.} 2016, \aap, 590,
  A76

\bibitem[{{Di Gesu} {et~al.}(2015){Di Gesu}, {Costantini}, {Ebrero},
  {Mehdipour}, {Kaastra}, {Ursini}, {Petrucci}, {Cappi}, {Kriss}, {Bianchi},
  {Branduardi-Raymont}, {De Marco}, {De Rosa}, {Kaspi}, {Paltani}, {Pinto},
  {Ponti}, {Steenbrugge}, \& {Whewell}}]{DiGe15}
{Di Gesu}, L., {Costantini}, E., {Ebrero}, J., {et~al.} 2015, \aap, 579, A42

\bibitem[{{Enya} {et~al.}(2002){Enya}, {Yoshii}, {Kobayashi}, {Minezaki},
  {Suganuma}, {Tomita}, \& {Peterson}}]{enya02}
{Enya}, K., {Yoshii}, Y., {Kobayashi}, Y., {et~al.} 2002, \apjs, 141, 45

\bibitem[{{Event Horizon Telescope Collaboration} {et~al.}(2019){Event Horizon
  Telescope Collaboration}, {Akiyama}, {Alberdi}, {Alef}, {Asada}, {Azulay},
  {Baczko}, {Ball}, {Balokovi{\'c}}, {Barrett}, {Bintley}, {Blackburn},
  {Boland}, {Bouman}, {Bower}, {Bremer}, {Brinkerink}, {Brissenden}, {Britzen},
  {Broderick}, {Broguiere}, {Bronzwaer}, {Byun}, {Carlstrom}, {Chael}, {Chan},
  {Chatterjee}, {Chatterjee}, {Chen}, {Chen}, {Cho}, {Christian}, {Conway},
  {Cordes}, {Crew}, {Cui}, {Davelaar}, {De Laurentis}, {Deane}, {Dempsey},
  {Desvignes}, {Dexter}, {Doeleman}, {Eatough}, {Falcke}, {Fish}, {Fomalont},
  {Fraga-Encinas}, {Friberg}, {Fromm}, {G{\'o}mez}, {Galison}, {Gammie},
  {Garc{\'\i}a}, {Gentaz}, {Georgiev}, {Goddi}, {Gold}, {Gu}, {Gurwell},
  {Hada}, {Hecht}, {Hesper}, {Ho}, {Ho}, {Honma}, {Huang}, {Huang}, {Hughes},
  {Ikeda}, {Inoue}, {Issaoun}, {James}, {Jannuzi}, {Janssen}, {Jeter}, {Jiang},
  {Johnson}, {Jorstad}, {Jung}, {Karami}, {Karuppusamy}, {Kawashima},
  {Keating}, {Kettenis}, {Kim}, {Kim}, {Kim}, {Kino}, {Koay}, {Koch}, {Koyama},
  {Kramer}, {Kramer}, {Krichbaum}, {Kuo}, {Lauer}, {Lee}, {Li}, {Li},
  {Lindqvist}, {Liu}, {Liuzzo}, {Lo}, {Lobanov}, {Loinard}, {Lonsdale}, {Lu},
  {MacDonald}, {Mao}, {Markoff}, {Marrone}, {Marscher}, {Mart{\'\i}-Vidal},
  {Matsushita}, {Matthews}, {Medeiros}, {Menten}, {Mizuno}, {Mizuno}, {Moran},
  {Moriyama}, {Moscibrodzka}, {Mul{\ensuremath{\ddot{}}}ler}, {Nagai}, {Nagar},
  {Nakamura}, {Narayan}, {Narayanan}, {Natarajan}, {Neri}, {Ni}, {Noutsos},
  {Okino}, {Olivares}, {Oyama}, {{\"O}zel}, {Palumbo}, {Patel}, {Pen}, {Pesce},
  {Pi{\'e}tu}, {Plambeck}, {PopStefanija}, {Porth}, {Prather},
  {Preciado-L{\'o}pez}, {Psaltis}, {Pu}, {Ramakrishnan}, {Rao}, {Rawlings},
  {Raymond}, {Rezzolla}, {Ripperda}, {Roelofs}, {Rogers}, {Ros}, {Rose},
  {Roshanineshat}, {Rottmann}, {Roy}, {Ruszczyk}, {Ryan}, {Rygl},
  {S{\'a}nchez}, {S{\'a}nchez-Arguelles}, {Sasada}, {Savolainen}, {Schloerb},
  {Schuster}, {Shao}, {Shen}, {Small}, {Sohn}, {SooHoo}, {Tazaki}, {Tiede},
  {Tilanus}, {Titus}, {Toma}, {Torne}, {Trent}, {Trippe}, {Tsuda}, {van
  Bemmel}, {van Langevelde}, {van Rossum}, {Wagner}, {Wardle}, {Weintroub},
  {Wex}, {Wharton}, {Wielgus}, {Wong}, {Wu}, {Young}, {Young}, {Younsi},
  {Yuan}, {Yuan}, {Zensus}, {Zhao}, {Zhao}, {Zhu}, {Anczarski}, {Baganoff},
  {Eckart}, {Farah}, {Haggard}, {Meyer-Zhao}, {Michalik}, {Nadolski},
  {Neilsen}, {Nishioka}, {Nowak}, {Pradel}, {Primiani}, {Souccar},
  {Vertatschitsch}, {Yamaguchi}, \& {Zhang}}]{M87}
{Event Horizon Telescope Collaboration}, {Akiyama}, K., {Alberdi}, A., {et~al.}
  2019, \apjl, 875, L5

\bibitem[{{Fabian} {et~al.}(2017){Fabian}, {Lohfink}, {Belmont}, {Malzac}, \&
  {Coppi}}]{Fabi17}
{Fabian}, A.~C., {Lohfink}, A., {Belmont}, R., {Malzac}, J., \& {Coppi}, P.
  2017, \mnras, 467, 2566

\bibitem[{{Fabian} {et~al.}(2015){Fabian}, {Lohfink}, {Kara}, {Parker},
  {Vasudevan}, \& {Reynolds}}]{Fabi15}
{Fabian}, A.~C., {Lohfink}, A., {Kara}, E., {et~al.} 2015, \mnras, 451, 4375

\bibitem[{{Faucher-Gigu{\`e}re} \& {Quataert}(2012)}]{Fauc12}
{Faucher-Gigu{\`e}re}, C.-A. \& {Quataert}, E. 2012, \mnras, 425, 605

\bibitem[{{Feruglio} {et~al.}(2015){Feruglio}, {Fiore}, {Carniani},
  {Piconcelli}, {Zappacosta}, {Bongiorno}, {Cicone}, {Maiolino}, {Marconi},
  {Menci}, {Puccetti}, \& {Veilleux}}]{Feru15}
{Feruglio}, C., {Fiore}, F., {Carniani}, S., {et~al.} 2015, \aap, 583, A99

\bibitem[{{Fiore} {et~al.}(2017){Fiore}, {Feruglio}, {Shankar}, {Bischetti},
  {Bongiorno}, {Brusa}, {Carniani}, {Cicone}, {Duras}, {Lamastra}, {Mainieri},
  {Marconi}, {Menci}, {Maiolino}, {Piconcelli}, {Vietri}, \&
  {Zappacosta}}]{Fiore17}
{Fiore}, F., {Feruglio}, C., {Shankar}, F., {et~al.} 2017, \aap, 601, A143

\bibitem[{{Garc{\'{\i}}a} {et~al.}(2014){Garc{\'{\i}}a}, {Dauser}, {Lohfink},
  {Kallman}, {Steiner}, {McClintock}, {Brenneman}, {Wilms}, {Eikmann},
  {Reynolds}, \& {Tombesi}}]{Garc14a}
{Garc{\'{\i}}a}, J., {Dauser}, T., {Lohfink}, A., {et~al.} 2014, \apj, 782, 76

\bibitem[{{Gaspari} \& {S{\k{a}}dowski}(2017)}]{Gaspari2017}
{Gaspari}, M. \& {S{\k{a}}dowski}, A. 2017, \apj, 837, 149

\bibitem[{{George} \& {Fabian}(1991)}]{George91}
{George}, I.~M. \& {Fabian}, A.~C. 1991, \mnras, 249, 352

\bibitem[{{Giacconi} {et~al.}(1979){Giacconi}, {Branduardi}, {Briel},
  {Epstein}, {Fabricant}, {Feigelson}, {Forman}, {Gorenstein}, {Grindlay},
  {Gursky}, {Harnden}, {Henry}, {Jones}, {Kellogg}, {Koch}, {Murray},
  {Schreier}, {Seward}, {Tananbaum}, {Topka}, {Van Speybroeck}, {Holt},
  {Becker}, {Boldt}, {Serlemitsos}, {Clark}, {Canizares}, {Markert}, {Novick},
  {Helfand}, \& {Long}}]{Giacconi79}
{Giacconi}, R., {Branduardi}, G., {Briel}, U., {et~al.} 1979, \apj, 230, 540

\bibitem[{{Gioia} {et~al.}(1990){Gioia}, {Maccacaro}, {Schild}, {Wolter},
  {Stocke}, {Morris}, \& {Henry}}]{Gioia90}
{Gioia}, I.~M., {Maccacaro}, T., {Schild}, R.~E., {et~al.} 1990, \apjs, 72, 567

\bibitem[{{Gofford} {et~al.}(2013){Gofford}, {Reeves}, {Tombesi}, {Braito},
  {Turner}, {Miller}, \& {Cappi}}]{Goff13}
{Gofford}, J., {Reeves}, J.~N., {Tombesi}, F., {et~al.} 2013, \mnras, 430, 60

\bibitem[{{Guainazzi} {et~al.}(1999){Guainazzi}, {Matt}, {Molendi}, {Orr},
  {Fiore}, {Grandi}, {Matteuzzi}, {Mineo}, {Perola}, {Parmar}, \&
  {Piro}}]{Guai99b}
{Guainazzi}, M., {Matt}, G., {Molendi}, S., {et~al.} 1999, \aap, 341, L27

\bibitem[{{Haardt} \& {Maraschi}(1991)}]{haar91}
{Haardt}, F. \& {Maraschi}, L. 1991, \apjl, 380, L51

\bibitem[{{Haardt} \& {Maraschi}(1993)}]{haar93}
{Haardt}, F. \& {Maraschi}, L. 1993, \apj, 413, 507

\bibitem[{{Halpern}(1984)}]{Halpern1984}
{Halpern}, J.~P. 1984, \apj, 281, 90

\bibitem[{{Harrison} {et~al.}(2013){Harrison}, {Craig}, {Christensen},
  {Hailey}, {Zhang}, {Boggs}, {Stern}, {Cook}, {Forster}, {Giommi},
  {Grefenstette}, {Kim}, {Kitaguchi}, {Koglin}, {Madsen}, {Mao}, {Miyasaka},
  {Mori}, {Perri}, {Pivovaroff}, {Puccetti}, {Rana}, {Westergaard}, {Willis},
  {Zoglauer}, {An}, {Bachetti}, {Barri{\`e}re}, {Bellm}, {Bhalerao},
  {Brejnholt}, {Fuerst}, {Liebe}, {Markwardt}, {Nynka}, {Vogel}, {Walton},
  {Wik}, {Alexander}, {Cominsky}, {Hornschemeier}, {Hornstrup}, {Kaspi},
  {Madejski}, {Matt}, {Molendi}, {Smith}, {Tomsick}, {Ajello}, {Ballantyne},
  {Balokovi{\'c}}, {Barret}, {Bauer}, {Blandford}, {Brandt}, {Brenneman},
  {Chiang}, {Chakrabarty}, {Chenevez}, {Comastri}, {Dufour}, {Elvis}, {Fabian},
  {Farrah}, {Fryer}, {Gotthelf}, {Grindlay}, {Helfand}, {Krivonos}, {Meier},
  {Miller}, {Natalucci}, {Ogle}, {Ofek}, {Ptak}, {Reynolds}, {Rigby},
  {Tagliaferri}, {Thorsett}, {Treister}, \& {Urry}}]{Harr13}
{Harrison}, F.~A., {Craig}, W.~W., {Christensen}, F.~E., {et~al.} 2013, \apj,
  770, 103

\bibitem[{{HI4PI Collaboration} {et~al.}(2016){HI4PI Collaboration}, {Ben
  Bekhti}, {Fl{\"o}er}, {Keller}, {Kerp}, {Lenz}, {Winkel}, {Bailin},
  {Calabretta}, {Dedes}, {Ford}, {Gibson}, {Haud}, {Janowiecki}, {Kalberla},
  {Lockman}, {McClure-Griffiths}, {Murphy}, {Nakanishi}, {Pisano}, \&
  {Staveley-Smith}}]{HI4PI}
{HI4PI Collaboration}, {Ben Bekhti}, N., {Fl{\"o}er}, L., {et~al.} 2016, \aap,
  594, A116

\bibitem[{{Jansen} {et~al.}(2001){Jansen}, {Lumb}, {Altieri}, {Clavel}, {Ehle},
  {Erd}, {Gabriel}, {Guainazzi}, {Gondoin}, {Much}, {Munoz}, {Santos},
  {Schartel}, {Texier}, \& {Vacanti}}]{Jans01}
{Jansen}, F., {Lumb}, D., {Altieri}, B., {et~al.} 2001, \aap, 365, L1

\bibitem[{{Kaastra} {et~al.}(2011){Kaastra}, {Petrucci}, {Cappi}, {Arav},
  {Behar}, {Bianchi}, {Bloom}, {Blustin}, {Branduardi-Raymont}, {Costantini},
  {Dadina}, {Detmers}, {Ebrero}, {Jonker}, {Klein}, {Kriss}, {Lubi{\'n}ski},
  {Malzac}, {Mehdipour}, {Paltani}, {Pinto}, {Ponti}, {Ratti}, {Smith},
  {Steenbrugge}, \& {de Vries}}]{Kaas11}
{Kaastra}, J.~S., {Petrucci}, P.-O., {Cappi}, M., {et~al.} 2011, \aap, 534, A36

\bibitem[{{Kallman} \& {Bautista}(2001)}]{Kall01}
{Kallman}, T. \& {Bautista}, M. 2001, \apjs, 133, 221

\bibitem[{Krolik \& Kriss(2001)}]{Krolik_2001}
Krolik, J.~H. \& Kriss, G.~A. 2001, The Astrophysical Journal, 561, 684

\bibitem[{{Laha} {et~al.}(2016){Laha}, {Guainazzi}, {Chakravorty}, {Dewangan},
  \& {Kembhavi}}]{Laha2016}
{Laha}, S., {Guainazzi}, M., {Chakravorty}, S., {Dewangan}, G.~C., \&
  {Kembhavi}, A.~K. 2016, \mnras, 457, 3896

\bibitem[{{Laha} {et~al.}(2014){Laha}, {Guainazzi}, {Dewangan}, {Chakravorty},
  \& {Kembhavi}}]{Laha2014}
{Laha}, S., {Guainazzi}, M., {Dewangan}, G.~C., {Chakravorty}, S., \&
  {Kembhavi}, A.~K. 2014, \mnras, 441, 2613

\bibitem[{{Lansbury} {et~al.}(2017){Lansbury}, {Alexander}, {Aird}, {Gandhi},
  {Stern}, {Koss}, {Lamperti}, {Ajello}, {Annuar}, {Assef}, {Ballantyne},
  {Balokovi{\'c}}, {Bauer}, {Brandt}, {Brightman}, {Chen}, {Civano},
  {Comastri}, {Del Moro}, {Fuentes}, {Harrison}, {Marchesi}, {Masini},
  {Mullaney}, {Ricci}, {Saez}, {Tomsick}, {Treister}, {Walton}, \&
  {Zappacosta}}]{Lans17}
{Lansbury}, G.~B., {Alexander}, D.~M., {Aird}, J., {et~al.} 2017, \apj, 846, 20

\bibitem[{{Longinotti} {et~al.}(2010){Longinotti}, {Costantini}, {Petrucci},
  {Boisson}, {Mouchet}, {Santos-Lleo}, {Matt}, {Ponti}, \&
  {Gon{\c{c}}alves}}]{Longinotti2010}
{Longinotti}, A.~L., {Costantini}, E., {Petrucci}, P.~O., {et~al.} 2010, \aap,
  510, A92

\bibitem[{{Mao} {et~al.}(2019){Mao}, {Mehdipour}, {Kaastra}, {Costantini},
  {Pinto}, {Branduardi-Raymont}, {Behar}, {Peretz}, {Bianchi}, {Kriss},
  {Ponti}, {De Marco}, {Petrucci}, {Di Gesu}, {Middei}, {Ebrero}, \&
  {Arav}}]{Mao2019}
{Mao}, J., {Mehdipour}, M., {Kaastra}, J.~S., {et~al.} 2019, \aap, 621, A99

\bibitem[{{Marinucci} {et~al.}(2018){Marinucci}, {Bianchi}, {Braito}, {Matt},
  {Nardini}, \& {Reeves}}]{Marinucci2018}
{Marinucci}, A., {Bianchi}, S., {Braito}, V., {et~al.} 2018, \mnras, 478, 5638

\bibitem[{{Markowitz} {et~al.}(2006){Markowitz}, {Reeves}, \&
  {Braito}}]{Markowitz2006}
{Markowitz}, A., {Reeves}, J.~N., \& {Braito}, V. 2006, \apj, 646, 783

\bibitem[{{Matt} {et~al.}(1991){Matt}, {Perola}, \& {Piro}}]{Matt91}
{Matt}, G., {Perola}, G.~C., \& {Piro}, L. 1991, \aap, 247, 25

\bibitem[{{Matzeu} {et~al.}(2019){Matzeu}, {Braito}, {Reeves}, {Severgnini},
  {Ballo}, {Caccianiga}, {Campana}, {Cicone}, {Della Ceca}, \&
  {Parker}}]{Matzeu19}
{Matzeu}, G.~A., {Braito}, V., {Reeves}, J.~N., {et~al.} 2019, \mnras, 483,
  2836

\bibitem[{{Matzeu} {et~al.}(2017){Matzeu}, {Reeves}, {Braito}, {Nardini},
  {McLaughlin}, {Lobban}, {Tombesi}, \& {Costa}}]{Matzeu17}
{Matzeu}, G.~A., {Reeves}, J.~N., {Braito}, V., {et~al.} 2017, \mnras, 472, L15

\bibitem[{{Middei} {et~al.}(2018{\natexlab{a}}){Middei}, {Bianchi}, {Cappi},
  {Petrucci}, {Ursini}, {Arav}, {Behar}, {Branduardi-Raymont}, {Costantini},
  {De Marco}, {Di Gesu}, {Ebrero}, {Kaastra}, {Kaspi}, {Kriss}, {Mao},
  {Mehdipour}, {Paltani}, {Peretz}, \& {Ponti}}]{Middei18}
{Middei}, R., {Bianchi}, S., {Cappi}, M., {et~al.} 2018{\natexlab{a}}, \aap,
  615, A163

\bibitem[{{Middei} {et~al.}(2018{\natexlab{b}}){Middei}, {Vagnetti}, {Tombesi},
  {Bianchi}, {Marinucci}, {Ursini}, \& {Tortosa}}]{Midd18b}
{Middei}, R., {Vagnetti}, F., {Tombesi}, F., {et~al.} 2018{\natexlab{b}}, \aap,
  618, A167

\bibitem[{{Miniutti} \& {Fabian}(2006)}]{Miniutti2006}
{Miniutti}, G. \& {Fabian}, A.~C. 2006, \mnras, 366, 115

\bibitem[{{Parker} {et~al.}(2018){Parker}, {Matzeu}, {Guainazzi},
  {Kalfountzou}, {Miniutti}, {Santos-Lle{\'o}}, \& {Schartel}}]{Parker18}
{Parker}, M.~L., {Matzeu}, G.~A., {Guainazzi}, M., {et~al.} 2018, \mnras, 480,
  2365

\bibitem[{{Parker} {et~al.}(2017){Parker}, {Pinto}, {Fabian}, {Lohfink},
  {Buisson}, {Alston}, {Kara}, {Cackett}, {Chiang}, {Dauser}, {De Marco},
  {Gallo}, {Garcia}, {Harrison}, {King}, {Middleton}, {Miller}, {Miniutti},
  {Reynolds}, {Uttley}, {Vasudevan}, {Walton}, {Wilkins}, \&
  {Zoghbi}}]{Parker17}
{Parker}, M.~L., {Pinto}, C., {Fabian}, A.~C., {et~al.} 2017, \nat, 543, 83

\bibitem[{{Piconcelli} {et~al.}(2004){Piconcelli}, {Jimenez-Bail{\'o}n},
  {Guainazzi}, {Schartel}, {Rodr{\'{\i}}guez-Pascual}, \&
  {Santos-Lle{\'o}}}]{Pico04}
{Piconcelli}, E., {Jimenez-Bail{\'o}n}, E., {Guainazzi}, M., {et~al.} 2004,
  \mnras, 351, 161

\bibitem[{{Piconcelli} {et~al.}(2005){Piconcelli}, {Jimenez-Bail{\'o}n},
  {Guainazzi}, {Schartel}, {Rodr{\'\i}guez-Pascual}, \&
  {Santos-Lle{\'o}}}]{Piconcelli2005}
{Piconcelli}, E., {Jimenez-Bail{\'o}n}, E., {Guainazzi}, M., {et~al.} 2005,
  \aap, 432, 15

\bibitem[{{Pinto} {et~al.}(2018){Pinto}, {Alston}, {Parker}, {Fabian}, {Gallo},
  {Buisson}, {Walton}, {Kara}, {Jiang}, {Lohfink}, \& {Reynolds}}]{Pinto2018}
{Pinto}, C., {Alston}, W., {Parker}, M.~L., {et~al.} 2018, \mnras, 476, 1021

\bibitem[{{Porquet} {et~al.}(2004){Porquet}, {Reeves}, {Uttley}, \&
  {Turner}}]{Porquet2004}
{Porquet}, D., {Reeves}, J.~N., {Uttley}, P., \& {Turner}, T.~J. 2004, \aap,
  427, 101

\bibitem[{{Reeves} {et~al.}(2018){Reeves}, {Braito}, {Nardini}, {Lobban},
  {Matzeu}, \& {Costa}}]{Reeves2018}
{Reeves}, J.~N., {Braito}, V., {Nardini}, E., {et~al.} 2018, \apjl, 854, L8

\bibitem[{{Reynolds}(1997)}]{Reyn97}
{Reynolds}, C.~S. 1997, \mnras, 286, 513

\bibitem[{{Rybicki} \& {Lightman}(1979)}]{Rybi79}
{Rybicki}, G.~B. \& {Lightman}, A.~P. 1979, {Radiative processes in
  astrophysics}

\bibitem[{{Serafinelli} {et~al.}(2019){Serafinelli}, {Tombesi}, {Vagnetti},
  {Piconcelli}, {Gaspari}, \& {Saturni}}]{Seraf19}
{Serafinelli}, R., {Tombesi}, F., {Vagnetti}, F., {et~al.} 2019, \aap, 627,
  A121

\bibitem[{{Sergeev} {et~al.}(2007){Sergeev}, {Klimanov}, {Chesnok}, \&
  {Pronik}}]{Serg07}
{Sergeev}, S.~G., {Klimanov}, S.~A., {Chesnok}, N.~G., \& {Pronik}, V.~I. 2007,
  Astronomy Letters, 33, 429

\bibitem[{{Smith} {et~al.}(2019){Smith}, {Tombesi}, {Veilleux}, {Lohfink}, \&
  {Luminari}}]{Smith2019}
{Smith}, R.~N., {Tombesi}, F., {Veilleux}, S., {Lohfink}, A.~M., \& {Luminari},
  A. 2019, arXiv e-prints, arXiv:1910.14583

\bibitem[{{Tombesi} \& {Cappi}(2014)}]{Tombesi2014}
{Tombesi}, F. \& {Cappi}, M. 2014, \mnras, 443, L104

\bibitem[{{Tombesi} {et~al.}(2012){Tombesi}, {Cappi}, {Reeves}, \&
  {Braito}}]{Tomb12}
{Tombesi}, F., {Cappi}, M., {Reeves}, J.~N., \& {Braito}, V. 2012, \mnras, 422,
  L1

\bibitem[{{Tombesi} {et~al.}(2013){Tombesi}, {Cappi}, {Reeves}, {Nemmen},
  {Braito}, {Gaspari}, \& {Reynolds}}]{Tomb13}
{Tombesi}, F., {Cappi}, M., {Reeves}, J.~N., {et~al.} 2013, \mnras, 430, 1102

\bibitem[{{Tombesi} {et~al.}(2011){Tombesi}, {Cappi}, {Reeves}, {Palumbo},
  {Braito}, \& {Dadina}}]{Tomb11}
{Tombesi}, F., {Cappi}, M., {Reeves}, J.~N., {et~al.} 2011, \apj, 742, 44

\bibitem[{{Tombesi} {et~al.}(2010){Tombesi}, {Cappi}, {Reeves}, {Palumbo},
  {Yaqoob}, {Braito}, \& {Dadina}}]{Tomb10}
{Tombesi}, F., {Cappi}, M., {Reeves}, J.~N., {et~al.} 2010, \aap, 521, A57

\bibitem[{{Tombesi} {et~al.}(2015){Tombesi}, {Mel{\'e}ndez}, {Veilleux},
  {Reeves}, {Gonz{\'a}lez-Alfonso}, \& {Reynolds}}]{Tomb15}
{Tombesi}, F., {Mel{\'e}ndez}, M., {Veilleux}, S., {et~al.} 2015, \nat, 519,
  436

\bibitem[{{Tombesi} {et~al.}(2014){Tombesi}, {Tazaki}, {Mushotzky}, {Ueda},
  {Cappi}, {Gofford}, {Reeves}, \& {Guainazzi}}]{Tomb14}
{Tombesi}, F., {Tazaki}, F., {Mushotzky}, R.~F., {et~al.} 2014, \mnras, 443,
  2154

\bibitem[{{Tombesi} {et~al.}(2017){Tombesi}, {Veilleux}, {Mel{\'e}ndez},
  {Lohfink}, {Reeves}, {Piconcelli}, {Fiore}, \& {Feruglio}}]{Tomb17}
{Tombesi}, F., {Veilleux}, S., {Mel{\'e}ndez}, M., {et~al.} 2017, \apj, 850,
  151

\bibitem[{{Tortosa} {et~al.}(2018){Tortosa}, {Bianchi}, {Marinucci}, {Matt}, \&
  {Petrucci}}]{Tort18}
{Tortosa}, A., {Bianchi}, S., {Marinucci}, A., {Matt}, G., \& {Petrucci}, P.~O.
  2018, \aap, 614, A37

\bibitem[{{Ursini} {et~al.}(2015){Ursini}, {Boissay}, {Petrucci}, {Matt},
  {Cappi}, {Bianchi}, {Kaastra}, {Harrison}, {Walton}, {di Gesu}, {Costantini},
  {De Marco}, {Kriss}, {Mehdipour}, {Paltani}, {Peterson}, {Ponti}, \&
  {Steenbrugge}}]{Ursi15}
{Ursini}, F., {Boissay}, R., {Petrucci}, P.-O., {et~al.} 2015, \aap, 577, A38

\bibitem[{{Veilleux} {et~al.}(2017){Veilleux}, {Bolatto}, {Tombesi},
  {Mel{\'e}ndez}, {Sturm}, {Gonz{\'a}lez-Alfonso}, {Fischer}, \&
  {Rupke}}]{Veil17}
{Veilleux}, S., {Bolatto}, A., {Tombesi}, F., {et~al.} 2017, \apj, 843, 18

\bibitem[{{Zubovas} \& {King}(2012)}]{Zubo12}
{Zubovas}, K. \& {King}, A. 2012, \apjl, 745, L34

\end{thebibliography}

\end{document}